\documentclass[12pt,onecolumn,draftclsnofoot]{IEEEtran}
\IEEEoverridecommandlockouts
\usepackage{stfloats}
\usepackage{cite}
\usepackage{amsmath,amssymb,amsfonts}
\usepackage{upgreek}
\usepackage{graphicx}
\usepackage{subfigure}
\usepackage{textcomp}
\usepackage{xcolor}
\usepackage{algorithm}
\usepackage[noend]{algpseudocode}
\usepackage{amsmath}
\usepackage[caption=false,font=normalsize,labelfont=sf,textfont=sf]{subfig}
\usepackage[linewidth=1pt]{mdframed}
\usepackage[mathscr]{eucal}
\usepackage{epstopdf}
\usepackage{cases}
\usepackage{xcolor}
\usepackage{bbding}
\usepackage{cleveref}
\usepackage{multicol}       
\usepackage{multirow}       
\usepackage{array}          
\usepackage{makecell}
\usepackage{flushend}
\usepackage{booktabs}
\definecolor{crimson}{RGB}{192,0,0}         
\definecolor{navy}{RGB}{47,85,151}         
\usepackage{colortbl}
\newtheorem{theorem}{\bf Theorem}
\newtheorem{corollary}{\bf Corollary}
\def\BibTeX{{\rm B\kern-.05em{\sc i\kern-.025em b}\kern-.08em
    T\kern-.1667em\lower.7ex\hbox{E}\kern-.125emX}}

\begin{document}
\title{Joint Cooperative Clustering and Power Control for Energy-Efficient Cell-Free XL-MIMO with Multi-Agent Reinforcement Learning}
\author{{Ziheng~Liu,~\IEEEmembership{Student Member,~IEEE}, Jiayi~Zhang,~\IEEEmembership{Senior Member,~IEEE}, Zhilong~Liu,~\IEEEmembership{Graduate Student Member,~IEEE}, Derrick~Wing~Kwan~Ng,~\IEEEmembership{Fellow,~IEEE}, and Bo~Ai,~\IEEEmembership{Fellow,~IEEE}}
\thanks{Z. Liu, J. Zhang, Z. Liu, and B. Ai are with the School of Electronic and Information Engineering and also with the Frontiers Science Center for Smart High-Speed Railway System, Beijing Jiaotong University, Beijing 100044, China (e-mail: \{zihengliu, zhangjiayi, zhilongliu, boai\}@bjtu.edu.cn).}
\thanks{D. W. K. Ng is with the School of Electrical Engineering and Telecommunications, University of New South Wales, NSW 2052, Australia (e-mail: w.k.ng@unsw.edu.au).}}
\maketitle
\begin{abstract}
In this paper, we investigate the amalgamation of cell-free (CF) and extremely large-scale multiple-input multiple-output (XL-MIMO) technologies, referred to as a CF XL-MIMO, as a promising advancement for enabling future mobile networks. To address the computational complexity and communication power consumption associated with conventional centralized optimization, we focus on user-centric dynamic networks in which each user is served by an adaptive subset of access points (AP) rather than all of them. We begin our research by analyzing a joint resource allocation problem for energy-efficient CF XL-MIMO systems, encompassing cooperative clustering and power control design, where all clusters are adaptively adjustable. Then, we propose an innovative double-layer multi-agent reinforcement learning (MARL)-based scheme, which offers an effective strategy to tackle the challenges of high-dimensional signal processing. In the section of numerical results, we compare various algorithms with different network architectures. These comparisons reveal that the proposed MARL-based cooperative architecture can effectively strike a balance between system performance and communication overhead, thereby improving energy efficiency performance.
It is important to note that increasing the number of user equipments participating in information sharing can effectively enhance SE performance, which also leads to an increase in power consumption, resulting in a non-trivial trade-off between the number of participants and EE performance.
\end{abstract}
\begin{IEEEkeywords}
Cooperative clustering, energy efficiency, multi-agent reinforcement learning, power control, XL-MIMO.
\end{IEEEkeywords}

\IEEEpeerreviewmaketitle
\section{Introduction}
With the ever-increasing number of communication devices, the escalating requirements for wireless communication have prompted academia to broaden their considerations, not only focusing on ``how fast" information can be transferred, in terms of spectral efficiency (SE), but also ``how green" the networks can become, in terms of energy efficiency (EE) \cite{[54]}.
Significant research efforts have been dedicated to developing novel wireless technologies to achieve the key performance indicators (KPIs) of beyond fifth-generation (B5G) and sixth-generation (6G) mobile communication networks, with the aim of realizing the vision of the Internet-of-Everything. Prominent examples of such technologies include cell-free (CF) \cite{[9],[10]} and extremely large-scale multiple-input multiple-output (XL-MIMO) \cite{[15],[16]}, which have garnered significant attention due to their potential in offering extensive degrees of freedom for spatial multiplexing and reliable massive access, leading to enhanced network performance.

As a remarkable technology, the core idea of CF massive MIMO (mMIMO) is to deploy numerous access points (APs) randomly throughout the coverage area, thereby shortening the communication distance between wirless transceivers \cite{[55]}. This approach is beneficial for enhancing system throughput and achieving massive access \cite{[8],[17]}. Unlike conventional cellular mMIMO systems that have defined cell boundaries, CF mMIMO operates without them and embraces a user-centric concept. This helps to eliminate inter-cell interference and overcome the performance limitations associated with small-cells \cite{[10]}.
On the other hand, from the perspective of electromagnetic (EM) characteristics, the promising XL-MIMO technology focuses on continuously expanding spatial degrees of freedom by increasing the number of antennas to enhance throughput, EE performance, etc. Meanwhile, as the EM operating region transitions from far-field to near-field, several new features have emerged, such as non-stationary effects and spherical wave propagating characteristics \cite{[20]}.
Specifically, in conventional mMIMO systems, the Rayleigh distance that separates the far-field and near-field regions is often deemed negligible. This assumption implies that all receivers are located in the far-field region and the EM wave can be approximately characterized by a planar wave \cite{[18],[21]}. However, with the significant increases in the number of antennas and carrier frequency, the near-field communication region is expected to expand dramatically, especially in XL-MIMO systems. This change renders it impractical to disregard the Rayleigh distance, positioning near-field communication as a pivotal scenario for next-generation mobile networks. As a result, the EM wave should be modeled adopting spherical wave models, contrasting starkly with the far-field models commonly found in current 5G mobile networks \cite{[20]}.
\subsection{Related Work}
In mMIMO systems impaired by severe inter-user interference, adaptive power control has proven advantageous in suppressing interference and optimizing system performance \cite{[45]}. As such, various schemes have been developed to address these issues \cite{[43],[44],[11]}.
The designs of these schemes are typically formulated as non-convex and NP-hard power control optimization problems \cite{[43]} and then handled by iterative \cite{[44]} or supervised learning algorithms \cite{[11]}. However, conventional centralized iterative optimization-based schemes, which have been extensively explored over the past few decades, tend to prioritize superior SE performance at the expense of elevated computational complexity, hindering the practical implementation of mMIMO systems. On the other hand, most of the existing model-free machine learning-based schemes can significantly reduce computational complexity, but they predominantly hinge on supervised learning. This poses an emerging challenge in obtaining prior optimal data as labels for training, especially in large-scale systems. Therefore, it is necessary to apply advanced optimization methods to address the aforementioned challenges.
Furthermore, channel hardening commonly occurs in mMIMO systems, where the randomly time-varying channel gains behave deterministically. Consequently, the optimization of transmission power can be based solely on large-scale fading information and spatial correlation, bypassing the need to account for small-scale fading information \cite{[30]}.

On a different note, multi-agent reinforcement learning (MARL) has emerged as a mature decision-making architecture for multi-agent scenarios, demonstrating excellent performance in terms of generalization and timeliness \cite{[34],[35]}. The goal of MARL is to enable a group of agents to collaboratively discover a globally optimal policy that maximizes the sum of their individual rewards.
In particular, recent works have adopted numerous effective MARL algorithms, such as the multi-agent deep deterministic policy gradient (MADDPG) \cite{[26],[27],[38]} and the multi-agent proximal policy optimization (MAPPO) \cite{[40]}, to tackle intractable problems in mMIMO systems due to their broad applicability, e.g. \cite{[22],[26],[27]}.
The authors in \cite{[26]} proposed two signal processing architectures, which utilized the combination of MARL and fuzzy logic to optimize the power control strategy, thereby maximizing the SE performance.
In addition, the authors in \cite{[27]} proposed two approaches, including single-agent RL and MARL, for optimizing user association and content caching jointly. However, while these MARL-based resource allocation schemes can achieve satisfactory performance, they require each agent to interact with all APs to train the value NNs, which limits the scalability of the designed schemes to accommodate varying numbers of APs.
To address the scalability challenges, the vast majority of existing papers focused on adopting AP clustering as a promising technology in mMIMO systems, ensuring that all active user equipments (UEs) are appropriately associated with specific APs to construct user-centric networks \cite{[38],[39]}.
For instance, the authors in \cite{[38]} developed a MARL framework for AP clustering in an environment with mobile UEs ensuring that each UE is served by the best possible user-centric personalized clustering of nearby APs.
However, it is worth noting that the MARL-based schemes mentioned above primarily focus on the conventional centralized training and decentralized execution (CTDE) architecture, which requires a training center to train the value NNs. This reliance on a training center introduces excessive computational complexity, limiting the practical implementation of mMIMO systems.

Moreover, the majority of the above works on mMIMO systems adopting MARL have operated under the overly idealistic and simplifying assumption of independent observability, where each agent can only access its own observations rather than those of others.
However, it has been proved that the ``partial observability" inherent in any actual environment may significantly influence the behavioral policies and performance of all agents \cite{[49]}. In this scenario, each agent can exchange its observed information and share decision-making policies with neighboring agents to compensate for its limited knowledge of multi-agent environments, and the returns from the actions taken by each agent depends on their joint decisions with other agents. This setup enables all agents to communicate more effectively to accomplish tasks.
To the best of the authors' knowledge, there is no research considering partial observability in mMIMO systems so far.
On the other hand, studies on conventional MARL systems incorporating partial observability mainly focused on the classic communication schemes (i.e., user grouping) to dynamically adjust the quantity and range of communication, aiming at enhancing collaboration and reducing communication power consumption.
Specifically, the modularity-based UE grouping algorithm (MUEGA) in \cite{[32]} and K-Means algorithm in \cite{[33]} are examples of such communication schemes. In particular, they group UEs with similar characteristics based on observed states, allowing mobile UEs to select relevant neighboring UEs for information sharing.
In practice, the existing communication schemes designed based on the aforementioned user grouping belongs to the non-overlapped mode, where each UE can only communicate with other UEs within their own group to supplement their local observations. However, this mode can result in redundant communication among UEs, which does not effectively reduce communication power consumption. Furthermore, due to the need for mobile UEs to communicate with other UEs through APs in mMIMO networks, merely merging the two independent schemes of AP clustering and communication clustering cannot meet the premise of information sharing among the UEs. As a remedy, all mobile UEs need to rely on APs as an intermediary communication medium.
\vspace{-0.3cm}
\subsection{Motivations and Contributions}
\begin{table*}[t]
  \centering
  \fontsize{9}{7}\selectfont
  \caption{Comparison of Relevant Research With This Paper.}
  \label{Paper_comparison}
    \begin{tabular}{ !{\vrule width1.2pt}  m{1.4 cm}<{\centering} !{\vrule width1.2pt}   m{1.4 cm}<{\centering} !{\vrule width1.2pt}  m{1.1 cm}<{\centering} !{\vrule width1.2pt}  m{2.1 cm}<{\centering}  !{\vrule width1.2pt} m{1.5 cm}<{\centering} !{\vrule width1.2pt} m{1.75 cm}<{\centering} !{\vrule width1.2pt} m{1.55 cm}<{\centering} !{\vrule width1.2pt}m{1.85 cm}<{\centering} !{\vrule width1.2pt}}
    \Xhline{1.2pt}
        \rowcolor{gray!30} \bf Ref. & \bf MARL Network & \bf Power Control & \bf Cooperation Clustering  &  \bf Scalability  &  \bf Decentralized & \bf Overlapped Mode & \bf Partial Observability \cr
    \Xhline{1.2pt}
        \bf Conf. \cite{[26]}   & \makecell[c]{\Checkmark} & \makecell[c]{\Checkmark} & \makecell[c]{\XSolidBrush} & \makecell[c]{\XSolidBrush} & \makecell[c]{\Checkmark} & \makecell[c]{\Checkmark} & \makecell[c]{\XSolidBrush}\cr\hline
        \cite{[44],[11]}   & \makecell[c]{\XSolidBrush} & \makecell[c]{\Checkmark} & \makecell[c]{\XSolidBrush} & \makecell[c]{\XSolidBrush} & \makecell[c]{\XSolidBrush} & \makecell[c]{\XSolidBrush} & \makecell[c]{\XSolidBrush}\cr\hline
         \cite{[38],[39]}   & \makecell[c]{\Checkmark} & \makecell[c]{\XSolidBrush} & \makecell[c]{\XSolidBrush} & \makecell[c]{\Checkmark} & \makecell[c]{\Checkmark} & \makecell[c]{\Checkmark} & \makecell[c]{\XSolidBrush}\cr\hline
          \cite{[32],[33]} & \makecell[c]{\XSolidBrush} & \makecell[c]{\XSolidBrush} & \makecell[c]{\Checkmark} & \makecell[c]{\XSolidBrush} & \makecell[c]{\XSolidBrush} & \makecell[c]{\XSolidBrush} & \makecell[c]{\XSolidBrush} \cr\hline
          \cite{[49]} & \makecell[c]{\Checkmark} & \makecell[c]{\XSolidBrush} & \makecell[c]{\Checkmark} & \makecell[c]{\XSolidBrush} & \makecell[c]{\XSolidBrush} & \makecell[c]{\Checkmark} & \makecell[c]{\Checkmark} \cr\hline
         \bf Proposed & \makecell[c]{\Checkmark} & \makecell[c]{\Checkmark} & \makecell[c]{\Checkmark} & \makecell[c]{\Checkmark} & \makecell[c]{\Checkmark} & \makecell[c]{\Checkmark} & \makecell[c]{\Checkmark}\cr\hline
    \Xhline{1.2pt}
    \end{tabular}
\end{table*}
Motivated by the above observations, we initially establish the premise of communication among the UEs based on whether they are served by the same AP. Then, we study the AP clustering and communication clustering collectively, termed as cooperative clustering. In this approach, the original CTDE architecture is transformed into a cooperative architecture, where all policy and value NNs are stationed at the agents rather than at the CPU. This eliminates the need for centralized training centers while significantly reducing computational complexity and communication power consumption.

Additionally, this paper commences by laying out the foundational framework of CF XL-MIMO systems, which integrates CF mMIMO and XL-MIMO technologies and focuses on exploring the EE optimization problem.
The comparisons between our work and several aforementioned existing works are summarized in Table \uppercase\expandafter{\romannumeral1}.
The major contributions of this paper are listed as follows:
\begin{itemize}
\item We first investigate a novel multi-AP multi-UE channel model for CF XL-MIMO systems within the near-field communication. Then, we derive the achievable uplink spectral efficiency (SE) expression under user-centric dynamic networks and compute the closed-form SE and EE expressions when MR combining is adopted.

\item We propose a joint cooperative clustering and power control scheme that leverages a MARL network, where each AP cluster is adaptively adjusted based on available channel state information. Specifically, the first layer's cooperative clustering network is responsible for adaptively selecting suitable AP clusters and neighboring UEs served by the same AP, aiming to limit the amount of communication overhead. For the second layer's power control network, it is mainly responsible for allocating appropriate transmission power to enhance throughput.

\item Our results demonstrate that adopting a suitable communication scheme is beneficial for striking an excellent balance between system performance and communication overhead. Compared with conventional centralized optimization, the proposed joint scheme not only achieves a higher EE, but also enjoys a substantial SE.
\end{itemize}

The rest of this paper is organized as follows. In Section \uppercase\expandafter{\romannumeral2}, we
discuss a near-field channel model and the uplink data transmission. Section \uppercase\expandafter{\romannumeral3} introduces a joint cooperative clustering and power control model for maximizing EE. Then, in Section \uppercase\expandafter{\romannumeral4}, we propose a double-layer architecture, which combines the cooperative clustering network for and the power control network. In Section \uppercase\expandafter{\romannumeral5}, numerical results and performance analysis for the proposed scheme with the conventional schemes are provided. Finally, the major conclusions and future directions are drawn in Section \uppercase\expandafter{\romannumeral6}.
\newcounter{mytempeqncnt_1}

\emph{\textbf{{    Notation}}}: Column vectors and matrices are denoted by boldface lowercase letters $\bf{x}$ and boldface uppercase letters $\bf{X}$, respectively. $\mathbb{E\{\cdot\}}$, ${\text{tr}\{\cdot\}}$, and $\triangleq$ represent the expectation operator, trace operator, and definitions, respectively. $\left(\cdot\right)$\textsuperscript{\emph{$\ast$}}, $\left(\cdot\right)$\textsuperscript{\emph{T}}, and $\left(\cdot\right)$\textsuperscript{\emph{\textrm{H}}} represent conjugate, transpose, and conjugate transpose, respectively.
$\text{diag}(\mathbf{A}_1,\ldots,\mathbf{A}_n)$ and $\text{vec}(\mathbf{A})$ denote a block-diagonal matrix and the column vector formed by the stack of the columns of $\mathbf{A}$, respectively. $\otimes$ and $\odot$ denote the Kronecker products and the element-wise products, respectively. The $N \times N$ zero matrix and identity matrix are denoted by $\mathbf{0}_{N}$ and $\mathbf{I}_{N}$, respectively. $|\cdot|$, $\|\cdot\|$, and $\nabla$ are the determinant of a matrix, the Euclidean norm, and gradient, respectively. $\mathbb{B}^n$, $\mathbb{Z}^n$, $\mathbb{R}^n$, and $\mathbb{C}^n$ represent the n-dimensional spaces of binary, integer, real, and complex numbers, respectively. Finally, $x\sim{{\cal N}_\mathbb{C}}\left({0},\sigma^2\right)$ with zero mean and variance $\sigma^2$ is a circularly symmetric complex Gaussian distribution vector.
\section{System Model}
As illustrated in Fig. 1, we consider a CF XL-MIMO system consisting of $M$ APs and $K$ UEs arbitrarily distributed in a wide coverage area and denote $\mathcal{M} \triangleq \{1,2,\ldots,M\}$ and $\mathcal{K} \triangleq \{1,2,\ldots,K\}$. Every AP is connected to a central processing unit (CPU) possessing substantial processing capabilities via fronthaul links, aiming to reduce the computation overhead. Both the APs and UEs incorporate a planar extremely large-scale surface composed of patch antennas (XL-surface) \cite{[1],[2]}, which has been widely adopted in practical scenarios due to its compact size, low profile, and ease of fabrication, and includes key features such as wireless communication, signal transmission, signal reception, and frequency selectivity. Specifically, the numbers of patch antennas per AP and UE are $N_r = N_{V_r}N_{H_r}$ and $N_s = N_{V_s}N_{H_s}$, respectively, where $N_{V_r}$ and $N_{V_s}$ denote the number of patch antennas in the vertical direction for the APs and UEs, respectively, and $N_{H_r}$ and $N_{H_s}$ denote the number of patch antennas in the horizontal direction for the APs and UEs, respectively. Additionally, we assume that the patch antenna spacing $\Delta_r$ and $\Delta_s$ at each AP and UE, respectively, are both less than half of the carrier wavelength $\lambda$, where densely packed sub-wavelength patch antennas with a particular size are incorporated. Then, we denote the overall vertical and horizontal length of the planar XL-surface at each AP by $L_{r,y} = N_{V_r}\Delta_r$ and $L_{r,x}=N_{H_r}\Delta_r$, respectively. Moreover, the patch antennas can be indexed row-by-row by $b \in [1,N_r]$ and the Cartesian coordinate of the $b$-th patch antenna at AP $m \in \mathcal{M}$ with respect to the origin is $\mathbf{r}_m^{b} = [r_{m,x}^{b},r_{m,y}^{b},r_{m,z}^{b}]^T$, where $r_{m,x}^{b}$, $r_{m,y}^{b}$, and $r_{m,z}^{b}$ are the coordinates of the x-axis, y-axis, and z-axis in the three-dimensional Cartesian coordinate system, respectively. Then, without loss of generality, the receive vector can be denoted as $\mathbf{a}_{r}(\mathbf{k},\mathbf{r})=[\mathbf{a}_{r,1}(\mathbf{k}_1,\mathbf{r}_1), \ldots, \mathbf{a}_{r,M}(\mathbf{k}_M,\mathbf{r}_M)] \in \mathbb{C}^{N_r \times M}$ such that the receive antenna response $\mathbf{a}_{r,m}(\mathbf{k}_m,\mathbf{r}_m) \in \mathbb{C}^{N_r}$ at AP $m$ is given by
\begin{figure}[t]
\centering
    \hspace{-0.3cm}
    \includegraphics[scale=0.145]{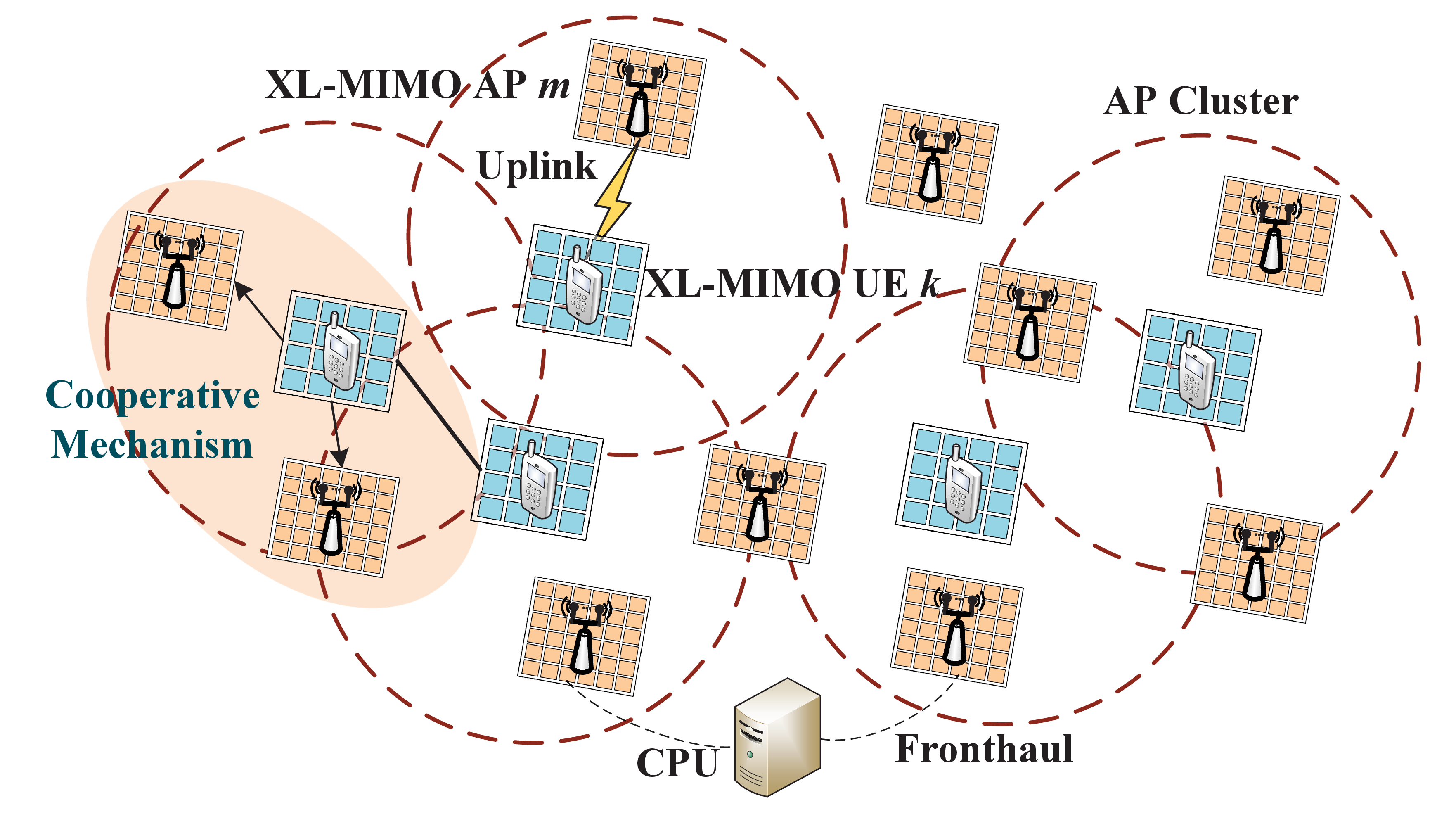}
    \caption{Illustration of a CF XL-MIMO system, where each red dotted circle represents the AP cluster formed by the central UE and its serving set of APs.
    \label{fig1}}
\end{figure}
\begin{equation}
\setcounter{equation}{1}
\mathbf{a}_{r,m}(\mathbf{k}_m,\mathbf{r}_m)=[e^{j\mathbf{k}_{m}(\varphi_m,\theta_m)^T{\mathbf{r}_m^{1}}}, \ldots, e^{j\mathbf{k}_{m}(\varphi_m,\theta_m)^T{\mathbf{r}_{m}^{N_r}}}]^T,
\label{eq1}
\end{equation}
where $\mathbf{k}_{m}(\varphi_m,\theta_m)=[k_{m,x},k_{m,y},k_{m,z}]=k[\cos(\theta_m)$ $\cos(\varphi_m),\cos(\theta_m)\sin(\varphi_m),\sin(\theta_m)] \in \mathbb{R}^3$ is the receive wave vector
with wavenumber $k=2\pi/\lambda$, the receive elevation angle $\theta_m$, and azimuth angle $\varphi_m$, respectively.

Similarly, the overall vertical and horizontal length of the planar XL-surface at each UE can be denoted as  $L_{s,y}=N_{V_s}\Delta_s$ and $L_{s,x}=N_{H_s}\Delta_s$, respectively. The patch antennas are indexed row-by-row by $u \in [1,N_s]$ and the Cartesian coordinate of the $u$-th patch antenna at UE $k \in \mathcal{K}$ with respect to the origin is $\mathbf{s}_k^{u} = [s_{k,x}^{u},s_{k,y}^{u},s_{k,z}^{u}]^T$, where $s_{k,x}^{u}$, $s_{k,y}^{u}$, and $s_{k,z}^{u}$ are the coordinates of the x-axis, y-axis, and z-axis in the three-dimensional Cartesian coordinate system, respectively. Then, the transmit vector can be denoted as $\mathbf{a}_{s}(\boldsymbol{\kappa},\mathbf{s})=[\mathbf{a}_{s,1}(\boldsymbol{\kappa}_1,\mathbf{s}_1), \ldots,\mathbf{a}_{s,K}(\boldsymbol{\kappa}_K,\mathbf{s}_K)] \in \mathbb{C}^{N_s \times K}$ and the transmit antenna response  $\mathbf{a}_{s,k}(\boldsymbol{\kappa}_k,\mathbf{s}_k) \in \mathbb{C}^{N_s}$ at UE $k$ with the elevation angle $\theta_k$ and azimuth angle $\varphi_k$ is given by
\begin{equation}
\setcounter{equation}{2}
\mathbf{a}_{s,k}(\boldsymbol{\kappa}_k,\mathbf{s}_k)=[e^{j\boldsymbol{\kappa}_{k}(\varphi_k,\theta_k)^T{\mathbf{s}_k^{1}}}, \ldots,e^{j\boldsymbol{\kappa}_{k}(\varphi_k,\theta_k)^T{\mathbf{s}_k^{N_s}}}]^T,
\label{eq2}
\end{equation}
where $\boldsymbol{\kappa}_{k}(\varphi_k,\theta_k)=[\kappa_{k,x},\kappa_{k,y},\kappa_{k,z}]=k[\cos(\theta_k)\cos(\varphi_k),$ $\cos(\theta_k)\sin(\varphi_k),\sin(\theta_k)] \in \mathbb{R}^3$ is the transmit wave vector.
\subsection{Channel Model}
In CF XL-MIMO systems, due to the associated large aperture, the near-field region is extended to a distance of kilometers such that it is more likely for UEs to be in the near-field. Therefore, the EM channel can be accurately described by the spherical wave model. Drawing parallel to the single-AP single-UE and single-AP multi-UE channel model proposed in \cite{[1],[2]},
these models were derived from Maxwell's equations, which serves as a foundational framework for describing EM fields.
The stochastic model was employed to portray non-line-of-sight EM propagation under arbitrary scattering conditions. Besides, due to the sparsity of equivalent angular channels in the wavenumber domain and the finite number of non-zero elements, spatial domain channels can be approximated by the finite sampling points within the effective bandwidth, i.e., the Fourier plane wave series expansion is non-zero only within the lattice ellipse \cite{[1],[2]}
\begin{equation}
\setcounter{equation}{3}
\begin{split}
\left \{
\begin{array}{ll}
\boldsymbol{\varepsilon}_r=\{(\ell_x,\ell_y)\in\mathbb{Z}^2:(\frac{\ell_x\lambda}{L_{r,x}})^2+(\frac{\ell_y\lambda}{L_{r,y}})^2\leqslant1\},\\
\boldsymbol{\varepsilon}_s=\{(m_x,m_y)\in\mathbb{Z}^2:(\frac{m_x\lambda}{L_{s,x}})^2+(\frac{m_y\lambda}{L_{s,y}})^2\leqslant1\},
\end{array}
\right.
\end{split}
\label{eq3}
\end{equation}
at the source and receiver, respectively. The cardinalities of the sets $\boldsymbol{\varepsilon}_r$ and $\boldsymbol{\varepsilon}_s$ can be denoted as $n_r = |\boldsymbol{\varepsilon}_r|$ and $n_s = |\boldsymbol{\varepsilon}_s|$, respectively.
Then, the overall small-scale fading channel $\mathbf{S}_{mk} \in \mathbb{C}^{N_r \times N_s}$ for the multi-AP multi-UE scenario can be derived as
\begin{equation}
\setcounter{equation}{4}
\mathbf{S}_{mk}=\sqrt{{N_r}{N_s}}\sum_{(\ell_x,\ell_y) \in \boldsymbol{\varepsilon}_r}\sum_{(m_x,m_y) \in \boldsymbol{\varepsilon}_s}S_a^{mk}(\ell_x,\ell_y,m_x,m_y)\mathbf{a}_{r,m}(\ell_x,\ell_y,\mathbf{r}_m)\mathbf{a}_{s,k}(m_x,m_y,\mathbf{s}_k),
\label{eq4}
\end{equation}
where $S_a^{mk}(\ell_x,\ell_y,m_x,m_y) \sim \mathcal{N}_\mathbb{C}(0,\sigma_{mk}^2(\ell_x,\ell_y,m_x,m_y))$ is the Fourier coefficient with variance $\sigma_{mk}^2(\ell_x,\ell_y,m_x,m_y)$,
and the receive signal of the $b$-th antenna $\mathbf{a}_{r,m}^{b}(\mathbf{k}_m,\mathbf{r}_m)$ at AP $m$ and the transmit signal of the $u$-th antenna $\mathbf{a}_{s,k}^{u}(\boldsymbol{\kappa}_k,\mathbf{s}_k)$ at UE $k$ can be denoted as
\begin{equation}
\setcounter{equation}{5}
\begin{split}
\left \{
\begin{array}{ll}
\mathbf{a}_{r,m}^{b}(\ell_x,\ell_y,\mathbf{r}_m) = \frac{1}{N_r}e^{-j\left(\frac{2\pi}{L_{r,x}}\ell_x r_{m,x}^{b} + \frac{2\pi}{L_{r,y}}\ell_y r_{m,y}^{b} + \sqrt{(\frac{2\pi}{\lambda})^2-\ell_x^2 -\ell_y^2} r_{m,z}^{b}\right)}, b = [1,\ldots,N_r],\\
\mathbf{a}_{s,k}^{u}(m_x,m_y,\mathbf{s}_k) = \frac{1}{N_s}e^{-j\left(\frac{2\pi}{L_{s,x}}m_x s_{k,x}^{u} + \frac{2\pi}{L_{s,y}}m_y s_{k,y}^{u} + \sqrt{(\frac{2\pi}{\lambda})^2-m_x^2 -m_y^2} s_{k,z}^{u}\right)}, u = [1,\ldots,N_s].
\end{array}
\right.
\end{split}
\label{eq5}
\end{equation}

Correspondingly, motivated by \cite{[1],[2]}, the small-scale fading channel model in (4) can be further rewritten as $\mathbf{S}_{mk}=\mathbf{U}_{r,m}\mathbf{S}_a^{mk}{\mathbf{U}_{s,k}}^H=\mathbf{U}_{r,m}(\mathbf{\Sigma}_{mk}\odot \mathbf{W}_{mk}){\mathbf{U}_{s,k}}^H \in \mathbb{C}^{N_r\times N_s}$, where $\mathbf{U}_{r,m} \in \mathbb{C}^{N_r\times n_r}$ and $\mathbf{U}_{s,k} \in \mathbb{C}^{N_s\times n_s}$ denote the matrices collecting all variances of $n_r$ and $n_s$ sampling points in $\mathbf{a}_{r,m}(\ell_x,\ell_y,\mathbf{r}_m)$ and $\mathbf{a}_{s,k}(m_x,m_y,\mathbf{s}_k)$, respectively. Specifically, since the columns of $\mathbf{U}_{r,m}$ and $\mathbf{U}_{s,k}$ describe the discretized receive and transmit plane-wave harmonics, which satisfy $\mathbf{U}_{r,m}{\mathbf{U}_{r,m}}^H = \mathbf{I}_{n_r}$ and $\mathbf{U}_{s,k}{\mathbf{U}_{s,k}}^H = \mathbf{I}_{n_s}$, respectively. Furthermore, $\mathbf{S}_a^{mk} = \boldsymbol{\Sigma}_{mk}\odot \mathbf{W}_{mk} \in \mathbb{C}^{n_r\times n_s}$ with $\mathbf{W}_{mk} \sim \mathcal{N}_\mathbb{C}(\mathbf{0}_{n_rn_s},\mathbf{I}_{n_rn_s})$ collects $\sqrt{N_rN_s}S_a^{mk}(\ell_x,\ell_y,m_x,m_y)$ for all $n_r\cdot n_s$ sampling points, and $\mathbf{\Sigma}_{mk}=(\boldsymbol{\sigma}_{r,m}\mathbf{1}_{n_s}^T)\odot(\mathbf{1}_{n_r}\boldsymbol{\sigma}_{s,k}^T)$, where $\boldsymbol{\sigma}_{r,m} \in \mathbb{R}^{n_r\times1}$ and $\boldsymbol{\sigma}_{s,k} \in \mathbb{R}^{n_s\times1}$ collect $\sqrt{N_r}\sigma_{r,m}(\ell_x,\ell_y)$ and $\sqrt{N_s}\sigma_{s,k}(m_x,m_y)$, respectively.

Moreover, based on the Cartesian coordinate of the $b$-th antenna $\mathbf{r}_{m}^{b}=[r_{m,x}^{b},r_{m,y}^{b},r_{m,z}^{b}]^T$ at AP $m$ and that of the $u$-th antenna $\mathbf{s}_{k}^{u}=[s_{k,x}^{u},s_{k,y}^{u},s_{k,y}^{u}]^T$ at UE $k$, the large-scale fading channel model can be denoted as $\mathbf{L}_{mk} \in \mathbb{C}^{N_r\times{N_s}}$ with $\mathbf{L}_{mk}^{b,u}=\sqrt{G_tF(\vartheta_{mk}^{b,u})}\lambda/({4\pi d_{mk}^{b,u}})$, where $d_{mk}^{b,u}=\|\mathbf{r}_{m}^{b}-\mathbf{s}_{k}^{u}\|$ and $\vartheta_{mk}^{b,u}$ are the distance and angle, respectively \cite{[4]}. Additionally, $G_t \in \mathbb{R}$ and $F(\vartheta_{mk}^{b,u}) \in \mathbb{R}$ denote the antenna gain and the normalized power radiation pattern, respectively, satisfying $\int_{0}^{\pi/2}G_tF(\vartheta_{mk}^{b,u})\sin\vartheta_{mk}^{b,u} d\vartheta_{mk}^{b,u} =1$. Then, we can derive the corresponding channel model based on the above analysis as $\mathbf{H}_{mk} = \mathbf{L}_{mk} \odot \mathbf{S}_{mk} \in \mathbb{C}^{N_r\times{N_s}}$ \cite{[5]}.
Generally, for a $16\times 16$-element planar XL-surface operating at a carrier frequency of 30 GHz, the distance $d_{mk}^{b,u}$ is very likely to be larger than the array aperture $D=16\times 0.5 \times 10^{-2}=0.08$ meters, i.e., $d_{mk}^{b,u} \gg D$, then it is reasonable to assume $d_{mk}^{b,u}\approx d_{mk}$
and $\mathbf{L}_{mk}^{b,u}\approx\beta_{mk}=\sqrt{G_tF(\vartheta_{mk})}\lambda/4\pi d_{mk}$ based on the Fresnel approximation \cite{[6]}.
Therefore, the corresponding channel model can be simplified as $\mathbf{H}_{mk,s}=\beta_{mk}\mathbf{S}_{mk} \in \mathbb{C}^{N_r\times{N_s}}$.
\subsection{Uplink Data Transmission}
During the uplink data transmission, all UEs simultaneously transmit their data symbols to all APs. Then, the received complex baseband signal $\mathbf{y}_m \in \mathbb{C}^{N_r}$ at AP $m$ is given by
\begin{equation}
\setcounter{equation}{6}
\mathbf{y}_m=\sum_{k \in \mathcal{K}}{\mathbf{H}_{mk}}\mathbf{d}_k+\mathbf{n}_m=\sum_{k \in \mathcal{K}}{\mathbf{H}_{mk}}\mathbf{P}_k\mathbf{x}_k+\mathbf{n}_m,
\label{eq6}
\end{equation}
where $\mathbf{n}_m \sim \mathcal{N}_\mathbb{C}(\mathbf{0}_{N_r},\sigma^2\mathbf{I}_{N_r})$ is the independent receiver noise with the noise power per antenna $\sigma^2$. Also, the transmitted signal composes of $\mathbf{d}_k=\mathbf{P}_k\mathbf{x}_k \in \mathbb{C}^{N_s}$ with the data symbol $\mathbf{x}_k \sim \mathcal{N}_\mathbb{C}(\mathbf{0}_{N_s},\mathbf{I}_{N_s})$ at UE $k$, where $\mathbf{P}_k \in \mathbb{C}^{{N_s}\times{N_s}}$ is the transmission power matrix, which should satisfy the power constraint as $\text{tr}(\mathbf{P}_k\mathbf{P}_k^H) \leqslant p_k$, where $p_k$ is the upper limit of the transmission power from UE $k$.
\section{Joint Model for Maximizing EE}
In this section, we first investigate the data transmission under user-centric networks. Then, we analyze the system performance and propose a joint cooperative clustering and power control model for maximizing the EE performance.
\subsection{User-centric Networks}
In a user-centric network, each UE chooses to communicate with a preferred set of neighbouring APs based on its unique requirements, where the UE is located in the middle of the circular coverage area and can be referred to as the central UE. Subsequently, the corresponding selected APs cooperate to jointly serve it. Based on this, we can denote that AP clustering $k$ is composed of central UE $k$ and the APs in the AP group (APG) $\mathcal{A}_k \subseteq \mathcal{M}$, as shown in Fig. 1.

In this paper, we consider a widely adopted overlapped clustering mode and define a unified large-scale coefficient threshold for all UEs $\beta_{c}$, as the criterion for structuring AP clustering. Specifically, it can be dynamically adjusted according to the environment to ensure that each UE can be served by the best possible user-centric personalized AP clustering. In particular, for the selection of a unified large-scale coefficient threshold $\beta_c$, it should be greater than the minimum large-scale coefficient within the system to ensure that each UE has a corresponding AP clustering.
Then, we define a set of indicator diagonal matrix $\mathbf{D}_{mk} \in \mathbb{B}^{{N_r}\times{N_r}}$, determining which APs communicate with which UEs, where $\mathbf{D}_{mk} = \mathbf{I}_{N_r}$ means that AP $m$ is associated to UE $k$ and $\mathbf{D}_{mk} = \mathbf{0}_{N_r}$, otherwise. As such, $\sum_{m \in \mathcal{M}}\mathbf{D}_{mk} = |\mathcal{A}_k|\mathbf{I}_{N_r}$, $\forall k \in \mathcal{K}$, holds. Then, let $\mathbf{V}_{mk} \in \mathbb{C}^{{N_r}\times{N_s}}$ denote the combining matrix designed by AP $m$ for UE $k$ and the effective receive combining matrix can be updated to
\begin{equation}
\setcounter{equation}{7}
\begin{split}
\mathbf{D}_{mk}\mathbf{V}_{mk}= \left \{
\begin{array}{ll}
\mathbf{V}_{mk}, \quad \text{if} \quad \beta_{mk} \geqslant \beta_{c} \rightarrow m \in \mathcal{A}_k, \\
\mathbf{0}_{N_r}, \quad \text{if} \quad \beta_{mk} < \beta_{c} \rightarrow m \notin \mathcal{A}_k.
\end{array}
\right.
\end{split}
\label{eq7}
\end{equation}

Then, the local estimation $\check{\mathbf{x}}_{mk} \in \mathbb{C}^{N_s}$ of the symbol $\mathbf{x}_k$ at AP $m$ can be denoted as
\begin{equation}
\setcounter{equation}{8}
\check{\mathbf{x}}_{mk}={\mathbf{V}_{mk}^H}{\mathbf{D}_{mk}^H}\mathbf{H}_{mk}\mathbf{P}_k\mathbf{x}_k+\sum_{l \in \mathcal{K}\backslash k}\mathbf{V}_{mk}^H{\mathbf{D}_{mk}^H}\mathbf{H}_{ml}\mathbf{P}_l\mathbf{x}_l+{\mathbf{V}_{mk}^H}{\mathbf{D}_{mk}^H}\mathbf{n}_m.
\label{eq8}
\end{equation}

Moreover, we notice that although the existing large-scale fading decoding method \cite{[9]} can achieve high SE performance among schemes that employing local combining at each AP, it requires a large amount of large-scale fading statistical parameters. These parameters grow quadratically with $M$, $K$, $N_r$, and $N_s$, which may be prohibitively large in CF XL-MIMO systems \cite{[3]}. For simplicity, the CPU can weight the local estimation $\{\check{\mathbf{x}}_{mk}: m \in \mathcal{M}\}$ by taking the average of the observations from all APs to obtain the final decoded symbol $\hat{\mathbf{x}}_{k}$ as
\begin{equation}
\setcounter{equation}{9}
\hat{\mathbf{x}}_{k}=\frac{1}{M}\Big(\sum_{m \in \mathcal{A}_k}{\mathbf{V}_{mk}^H}\mathbf{H}_{mk}\mathbf{P}_k\mathbf{x}_k+
\sum_{m \in \mathcal{A}_k}\sum_{l \in \mathcal{K}\backslash k}{\mathbf{V}_{mk}^H}\mathbf{H}_{ml}\mathbf{P}_l\mathbf{x}_l+ \sum_{m \in \mathcal{A}_k}{\mathbf{V}_{mk}^H}\mathbf{n}_m\Big).
\label{eq9}
\end{equation}

Accordingly, we can derive the uplink achievable SE based on local processing and simple centralized decoding as per the following theorem \cite{[3]}.
\begin{theorem}
\emph{An achievable SE of UE k in the considered CF XL-MIMO system is}
\begin{equation}
\setcounter{equation}{10}
\begin{aligned}
\text{SE}_k = \log_{2}{\left|\mathbf{I}_{N_s}+\mathbf{E}_k^H\mathbf{\Psi}_k^{-1}\mathbf{E}_k\right|},
\label{eq10}
\end{aligned}
\end{equation}
\emph{where} $\mathbf{E}_k \triangleq \sum_{m \in \mathcal{A}_k}\mathbb{E}\{{\mathbf{V}_{mk}^H}\mathbf{H}_{mk}\}\mathbf{P}_{k}$
\emph{,} $\mathbf{\Psi}_k \triangleq$ $\sum_{l \in \mathcal{K}}$ $\sum_{m \in \mathcal{A}_l}\sum_{m' \in \mathcal{A}_l}\mathbb{E}\{{\mathbf{V}_{mk}^H}\mathbf{H}_{ml}\mathbf{\bar{P}}_{l}\mathbf{H}_{m'l}^H\mathbf{V}_{m'k}\}-\mathbf{E}_k\mathbf{E}_k^H$
+ $\sum_{m \in \mathcal{A}_k}\mathbb{E}\{{\mathbf{V}_{mk}^H}\mathbf{n}_m\mathbf{n}_m^H\mathbf{V}_{mk}\}$ \emph{, and} $\mathbf{\bar{P}}_{l} \triangleq \mathbf{P}_l\mathbf{P}_l^H$.
\end{theorem}

To improve the performance of the SE, one can optimize the combining matrix $\mathbf{V}_{mk}$. For instance,
MR combining with $\mathbf{V}_{mk}=\mathbf{H}_{mk}$ or local minimum mean-squared error (L-MMSE) combining with $\mathbf{V}_{mk}=(\sum_{l \in \mathcal{K}}\mathbf{H}_{ml}\mathbf{\bar{P}}_{l}\mathbf{H}_{ml}^{H}+\sigma^2 \mathbf{I}_{N_r})^{-1}\mathbf{H}_{mk}\mathbf{P}_{k}$ are popular candidates for (10).
In practice, MR combining does not necessitate any matrix inversion that enjoys a low computational complexity and is more suitable for the practical implementation in CF XL-MIMO systems \cite{[8]}.
As a result, MR combining is considered in this paper and the corresponding closed-form SE expression is revealed in the following corollary.
\begin{corollary}
\emph{For MR combining $\mathbf{V}_{mk}=\mathbf{H}_{mk}$, we can derive the closed-form SE expression for UE $k$ as}
\begin{equation}
\setcounter{equation}{11}
\begin{aligned}
\text{SE}_{k,c} = \log_{2}{\left|\mathbf{I}_{N_s}+\mathbf{E}_{k,c}^H\mathbf{\Psi}_{k,c}^{-1}\mathbf{E}_{k,c}\right|},
\label{eq11}
\end{aligned}
\end{equation}
\emph{where} $\mathbf{E}_{k,c}=\sum_{m \in \mathcal{A}_k}\mathbf{Z}_{mk}\mathbf{P}_{k}$ \emph{with} $\mathbf{Z}_{mk}= \mathbb{E}\{\mathbf{H}_{mk}^H\mathbf{H}_{mk}\}$
\emph{and} $\mathbf{\Psi}_{k,c} =\sum_{l \in \mathcal{K}}$ $\sum_{m \in \mathcal{A}_l}\sum_{m' \in \mathcal{A}_l}\mathbf{T}_{kl}^{mm'}-\mathbf{E}_{k,c}\mathbf{E}_{k,c}^H$
+ $\sigma^2\sum_{m \in \mathcal{A}_k}\mathbf{Z}_{mk}$ \emph{with} $\mathbf{T}_{kl}^{mm'} = \mathbb{E}\{{\mathbf{H}_{mk}^H}\mathbf{H}_{ml}\mathbf{\bar{P}}_{l}\mathbf{H}_{m'l}^H\mathbf{H}_{m'k}\}$.
\end{corollary}
\begin{IEEEproof}
The proof follows a similar approach as [31, Theorem 3] and is omitted due to page limitations.
\end{IEEEproof}
\vspace{-0.3cm}
\subsection{Cooperative Clustering Model}
In terms of user-centric cooperative clustering, due to the presence of partial observability and non-stationary in the actual environment, each user can only access their own local observations. Therefore, they need to exchange their observed information (e.g. large-scale fading information or channel state information) and share decision-making strategies (e.g. assigned action information) with other UEs to supplement the lack of full knowledge about the environment. Specifically, each UE only selects neighboring UEs served by the same AP for information sharing, rather than all UEs.
For case of illustration, let $\mathbf{o}_k=[{o}_{k,1},\ldots,{o}_{k,k},\ldots,{o}_{k,k'},\ldots,{o}_{k,K}]^T \in \mathbb{B}^{{K}}$ with ${o}_{k,k} =1$ denote the binary indicator function between UE $k$ and other UEs $k' \in \mathcal{K} \backslash k$, where ${o}_{k,k'} =1$ means UE $k$ can share information with UE $k'$, and ${o}_{k,k'} = 0$, otherwise. According to the indicator diagonal matrix $\mathbf{D}_{mk} \in \mathbb{B}^{{N_r}\times{N_r}}$, the indicator function can be defined as
\begin{equation}
\setcounter{equation}{12}
\begin{split}
{o}_{k,k'}= \left \{
\begin{array}{ll}
1, \qquad \text{if} \quad \sum_{m \in \mathcal{M}}|D_{mk}^bD_{mk'}^b| \geqslant 1, \\
0, \qquad \text{if} \quad \sum_{m \in \mathcal{M}}|D_{mk}^bD_{mk'}^b| < 1.
\end{array}
\right.
\end{split}
\label{eq12}
\end{equation}
\subsection{Power Consumption Model and Energy Efficiency}
Without loss of generality, we adopt the power consumption model following \cite{[46]} for CF mMIMO systems and also extend it to user-centric CF XL-MIMO systems. The power consumption model captures the following main components: a) the radio site power consumption including the power consumed at the APs \{$P_m^\mathrm{ap}: \forall m$\}, the active UEs \{$P_k^\mathrm{ue}: \forall k$\}, communication connections \{$P_k^\mathrm{c}: \forall k$\}; b) the CPU power consumption $P_\mathrm{cpu}$; and c) the power consumption of signal processing $P_\mathrm{sp}$. Mathematically, the total power consumption $P_\mathrm{total}$ can be modeled by \cite{[46]}
\begin{equation}
\setcounter{equation}{13}
\begin{aligned}
P_\mathrm{total} = \sum_{m \in \mathcal{M}}P_m^\mathrm{ap} + \sum_{k \in \mathcal{K}}P_k^\mathrm{ue} + \sum_{k \in \mathcal{K}}P_k^\mathrm{con} + P_\mathrm{cpu}.
\label{eq13}
\end{aligned}
\end{equation}
We will now model each of these terms in details.

The power consumption related to AP $m \in {\mathcal{M}}$ is
\begin{equation}
\setcounter{equation}{14}
\begin{aligned}
P_m^\mathrm{ap} = N_rP_{m}^\mathrm{c,ap} + \sum_{k \in \mathcal{K}}|\mathbf{D}_{mk}|P_{m}^\mathrm{p,ap} + P_m^\mathrm{fh,ap},
\label{eq14}
\end{aligned}
\end{equation}
where $P_{m}^\mathrm{c,ap}$ is the internal circuit power per antenna, $P_{m}^\mathrm{p,ap}$ is the power consumption for processing received signals of the served UEs, and $P_m^\mathrm{fh,ap}$ is the power consumption adopted for the fronthaul connection between the CPU and AP $m$.

The power consumption at generic UE $k \in {\mathcal{K}}$ is
\begin{equation}
\setcounter{equation}{15}
\begin{aligned}
P_k^\mathrm{ue} = N_sP_{k}^\mathrm{c,ue} + \text{tr}(\mathbf{P}_k\mathbf{P}_k^H),
\label{eq15}
\end{aligned}
\end{equation}
where $P_{k}^\mathrm{c,ue}$ is the internal circuit power per antenna.

Information is exchanged among UEs via dedicated communication connections as a supplement to their own local information. Then, the communication power consumption of UE $k$ is composed of the number of links, time, and bandwidth, which can be expressed as
\begin{equation}
\setcounter{equation}{16}
\begin{aligned}
P_k^\mathrm{con} = N_s|\mathbf{o}_k|P_{k}^\mathrm{con,ue} + \frac{B\mathcal{C}_\mathrm{con}}{\mathcal{L}_\mathrm{sys}},
\label{eq16}
\end{aligned}
\end{equation}
where $P_{k}^\mathrm{con,ue}$ is the communication power per UE's patch antenna, $\mathcal{C}_\mathrm{con}$ is the computational complexity of information exchange, $B$ is the channel bandwidth, and $\mathcal{L}_\mathrm{sys}$ = 75 Gfops/W is the system computing efficiency, respectively.

The CPU is responsible for processing the data symbols of all UEs, with power consumption
\begin{equation}
\setcounter{equation}{17}
\begin{aligned}
P_\mathrm{cpu} = P_\mathrm{cpu}^\mathrm{fix} + \sum_{k \in \mathcal{K}}\mathrm{SE}_{k,c}(\mathcal{P}_\mathrm{cpu}^\mathrm{en} + \mathcal{P}_\mathrm{cpu}^\mathrm{de}) + \frac{B\mathcal{C}_\mathrm{cpu}}{\mathcal{L}_\mathrm{sys}},
\label{eq17}
\end{aligned}
\end{equation}
where $P_\mathrm{cpu}^\mathrm{fix}$ is the fixed power consumption of CPU, $\mathcal{P}_\mathrm{cpu}^\mathrm{en}$, $\mathcal{P}_\mathrm{cpu}^\mathrm{de}$ are the power consumption density for the final encoding and decoding at the CPU, respectively, and $\mathcal{C}_\mathrm{cpu}$ is the computational complexity of signal processing.

With the defined power consumption model and the closed-form SE expression, the total EE can be modeled as
\begin{equation}
\setcounter{equation}{18}
\begin{aligned}
{\text{EE}_{c}} = \frac{{\sum_{k \in \mathcal{K}}\text{SE}_{k,c}}}{P_\mathrm{total}}.
\label{eq18}
\end{aligned}
\end{equation}
\subsection{EE Maximization}
In this subsection, to unleash the potential of CF XL-MIMO systems and strike a balance between system performance and communication overhead, we focus on the joint design of cooperative clustering and power control to maximize EE performance.
By leveraging the properties of channel hardening \cite{[30]}, one can optimize the transmission power matrix $\mathbf{P}_{k}$, the indicator diagonal matrix $\mathbf{D}_{mk}$, and the indicator function $\mathbf{o}_{k}$ based on large-scale fading coefficients, bypassing the need to account the relatively fast time-varying small-scale fading coefficients. Then, with the practical constraints on transmission power and cooperative clustering in place, the joint optimization for maximizing EE can be mathematically formulated as
\begin{equation}
\setcounter{equation}{19}
\begin{aligned}
\max_{\{\mathbf{P}_k,\mathbf{D}_{mk},\mathbf{o}_k\}}  \text{EE}_{c}=\frac{{\sum_{k \in \mathcal{K}}\text{SE}_{k,c}}}{P_\mathrm{total}}&=\frac{{\sum_{k \in \mathcal{K}}\log_{2}{\left|\mathbf{I}_{N_s}+\mathbf{E}_{k,c}^H\mathbf{\Psi}_{k,c}^{-1}\mathbf{E}_{k,c}\right|}}}{P_\mathrm{total}}\\
\mbox{s.t.} \quad\qquad\qquad\thickspace\thickspace \text{tr}(\mathbf{P}_k\mathbf{P}_k^H) &\leqslant p_k, \forall k \in \mathcal{K},\\
\sum_{m \in \mathcal{M}}\mathbf{D}_{mk} &= |\mathcal{B}_k|\mathbf{I}_{N_r}, \forall k \in \mathcal{K},\\
\sum_{k' \in \mathcal{K}}o_{k,k'} &\leqslant 1 + \sum_{k' \in \mathcal{K}}\sum_{m \in \mathcal{M}}|D_{mk}^bD_{mk'}^b|, \forall k \in \mathcal{K}.\\
\label{eq19}
\end{aligned}
\end{equation}

It can be verified that the joint design problem in (19) is non-convex. Also, the computational complexity of applying conventional optimization-based schemes is prohibitively high \cite{[44]}, making them unsuitable for practical high-dimensional continuous control problems with partial observability and non-stationary. Therefore, in the next section, we introduce a novel double-layer MARL-based scheme that overcomes the aforementioned challenges.
\section{Proposed MADDPG-based Framework}
In this section, we propose a novel double-layer paradigm for large-scale MARL networks, called dynamic clustering-power (DCP)-MADDPG. It aims at maximizing the EE performance in mobile CF XL-MIMO systems by jointly optimizing the cooperative clustering and transmission power, with a focus on optimizing the uplink. Although there is no direct optimization for the downlink, the designed DCP-MADDPG scheme for the uplink is also applicable for the downlink due to channel reciprocity and duality.
\subsection{Markov Decision Process Model}
In common multi-agent networks, all goal-oriented agents obtain feedback through interaction with the environment to develop optimal strategies. In general, RL is studied utilizing Markov decision process (MDP) characterized by a tuple $<\mathcal{S}, \mathcal{A}, \mathcal{R}, \mathcal{P}, \gamma>$, where $\mathcal{S} = [s_{0},\ldots,s_{t},\ldots]$ with the observed state ${s}_{t}$ and $\mathcal{A} = [a_{0},\ldots,a_{t},\ldots]$ with the assigned action ${a}_{t}$ are state and action space, respectively. $\mathcal{R} = [r_{0},\ldots,r_{t},\ldots]$ is the comprehensive evaluation function consisting of rewards and penalties ${r}_{t}$ at $t$ time slot, where rewards are allocated to the positive value of ideal actions to encourage agents to take action to achieve specific goals, such as SE and EE. On the contrary, penalty is a negative value assigned to undesirable or suboptimal performance actions to prevent agents from taking actions that may have negative consequences or are inconsistent with the desired system behavior, such as communication overhead and computational complexity. Additionally, $\mathcal{P}:(\mathcal{S},\mathcal{A})\rightarrow\mathcal{S}$ and $\gamma$ are the state transition function and the discounted factor, respectively.

Recently, RL algorithms have been applied to handle numerous challenging resource allocation optimization problems in mobile systems, e.g. \cite{[26],[27]}. Indeed, numerous efficient algorithms have been derived, which can be broadly divided into three implementation forms: policy-based approaches, value-based approaches, and actor-critic approaches \cite{[34],[35]}. In the following, we briefly discuss these approaches.

\itshape \textbf{1) Policy-based approaches }\upshape construct a policy function $\mu({s_t})$, where the observed state $s_t$ directly maps the assigned action $a_t$ or the probability distribution of the assigned action $p({a_t})$, without the need to estimate the Q-value of the optimal strategy, making it suitable for scenarios with the continuous action space $\mathcal{A}_c$. However, conventional policy-based approaches often necessitate a complete state sequence and iteratively update the corresponding policy functions $\mu({s_t})$ separately, this makes it difficult to achieve convergence.

\itshape \textbf{2) Value-based approaches }\upshape construct and maintain a Q-table to acquire an optimal strategy for maximizing the expected return on all subsequent actions starting from the current state, where each item $Q(s_t, a_t)$ in the Q-table represents the expected return on an assigned action $a_t$ taken in the observed state $s_t$. Although such value-based approaches generally result in a high data utilization rate, they can only solve problems with small numbers of countable observed states and assigned actions due to the fact that each $(s_t, a_t)$ tuple needs to correspond to a Q-value, which has high computational complexity in large-scale scenarios. In addition, value-based approaches map the state-action to Q-values through deep neutron networks, making them effective in the continuous state space $\mathcal{S}_c$, such as DQN \cite{[40]}. However, due to the fact that the optimal value $Q(s_t, a_t)$ is unknown in actual scenarios, DQN is not suitable for the continuous action space $\mathcal{A}_c$, but only for scenarios with the discrete action space $\mathcal{A}_d$.

\itshape \textbf{3) Actor-Critic approaches }\upshape construct an actor network and a critic network, where the actor network exploits the policy function, which is responsible for assigning actions and interacting with the environment, while the critic network exploits the value function, which is responsible for evaluating the performance of the actor network and updating the strategy for the actor network.
Moreover, considering that the critic network fits the action-value function in the continuous space $\mathcal{A}_c$ and the actor network does not need to find the Q-value of the optimal strategy, it indicates that the actor-critic scheme is particularly appealing for continuous optimization space $\mathcal{S}_c$ and $\mathcal{A}_c$, such as DDPG \cite{[27],[38]}.

Based on the above analysis, considering the existence of continuous optimization space in joint problem, it is advisable to map the CF XL-MIMO environment into multi-agent systems and adopt the DDPG as the underlying architecture, which consists of a policy network $\mu({s_t}|\theta_{\pi})$ and a value network $Q({s_t,a_t}|\theta_{Q_\pi})$.
\subsection{Joint MARL-based Network Design}
\begin{figure*}[t]
\centering
    \includegraphics[scale=0.32]{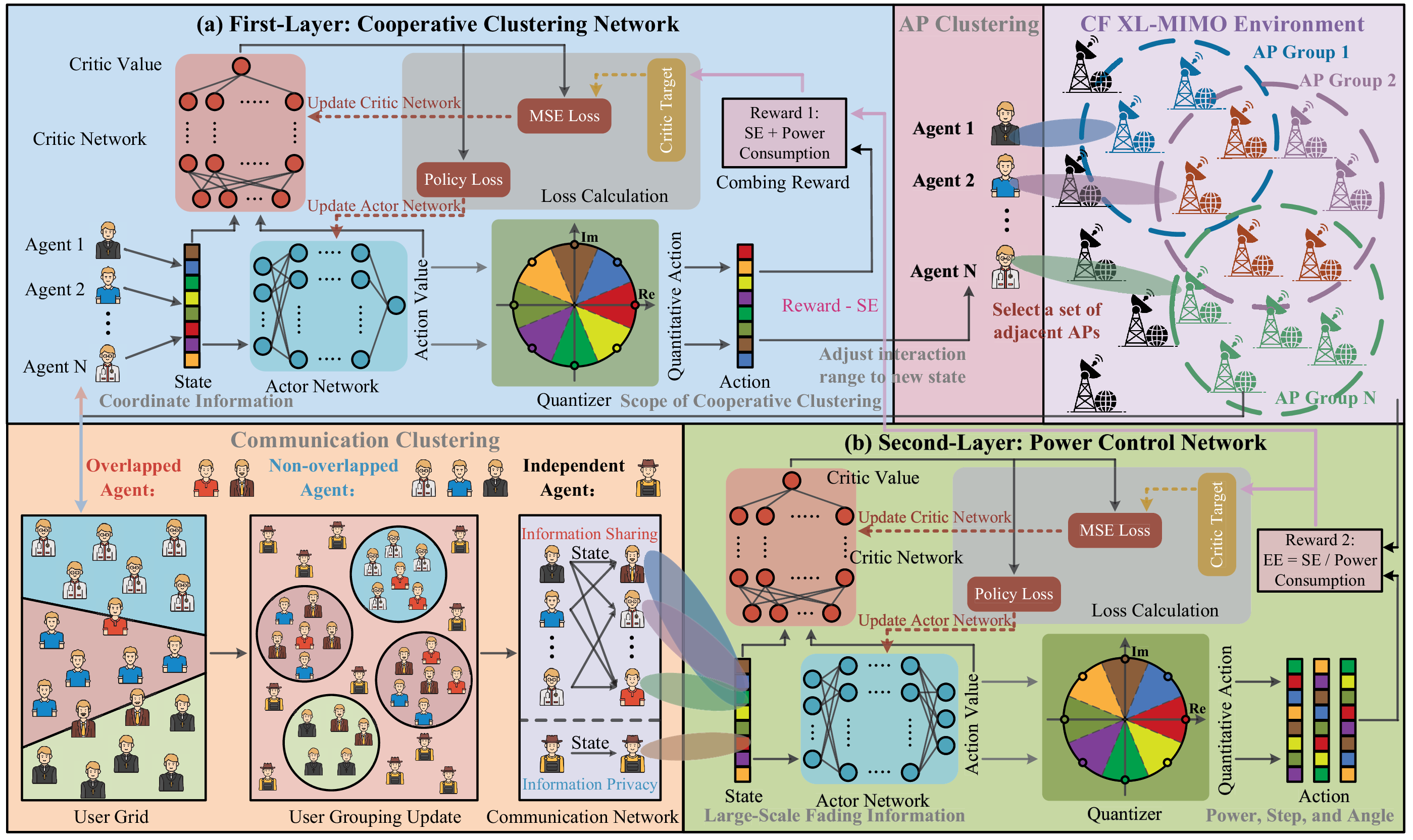}
    \caption{Illustration of a double-layer optimized architecture combining CF XL-MIMO system and MARL network, including the cooperative clustering network and the power control network. In this architecture, the CF XL-MIMO system provides state information inputs to the MARL network, while the MARL network updates all clusters and power coefficients in the CF XL-MIMO system through policy optimization.
    \label{fig1}}
\end{figure*}
In general, there are two architectures for addressing the joint cooperative clustering and power control problem in CF XL-MIMO systems, namely centralized and decentralized architectures \cite{[32],[33]}. For the former, it exploits the global information (i.e., each UE communicates with all remaining UEs to complement their local data) to enhance system performance, while the latter focuses on reducing communication power consumption without sacrificing too much performance.
Unfortunately, although these architectures may excel in a specific metric, they always impose a certain overly idealistic assumption of independent observability \cite{[32],[33],[26]}. It has been proven that partial observability that exists in any actual MARL environment may have a significant impact on the observed state and adopted polices of all agents  \cite{[49]}.
Moreover, for conventional cooperative schemes, i.e., the MUEGA, the K-Means, the variable communication quantity (Q-Variable), and the variable communication range (R-Variable), they force each UE to only communicate with other UEs within their own group to augment their local observations, resulting in redundant communication that is not conducive to reducing communication overhead. Therefore, we propose an extended version of the conventional MADDPG, namely DCP-MADDPG, which improves the network of MADDPG into a double-layer MARL network, as shown in Fig. 2.

In this double-layer network, we define the first layer as a cooperative clustering network, which dynamically selects suitable AP clusters and neighboring agents served by the same AP based on the observed coordinate information provided in CF XL-MIMO systems. This ensures that each agent can be served by the best possible user-centric clustering and supplement its local data by utilizing the observed information of neighboring agents. On the other hand, the second layer is defined as a power control network, which allocates the required power coefficient for each agent based on the large-scale fading information observed from the CF XL-MIMO system. Such a double-layer dynamic MARL network continuously updates the cluster selection and power allocation strategies in the CF XL-MIMO system through continuous collaboration between UEs to maximize system performance. Moreover, the cooperative clustering network designed in the first layer adopts a collaborative strategy in the initial AP selection scheme to further allocate APs outside the neighborhood to UEs within the system, thereby providing fair performance for each UE as much as possible by expanding the service domain.
\vspace{-0.1cm}
\subsection{First-Layer: Cooperative Clustering Network}
In the cooperative clustering network, we define all $K$ UE in mobile CF XL-MIMO systems as agents, while all antennas deployed by the same UE are considered as a whole for analysis. Considering that the first layer network maps the scope of cooperative clustering through the observed global position information, we can define the corresponding observed state and assigned action at $t$ time slot as $s_{t,(1)}=(d_{1},\ldots,d_{K})$ in conjunction with $d_{k}=\sum_{m \in \mathcal{M}}d_{mk}=\sum_{m \in \mathcal{M}}\|\mathbf{r}_{m}^{b}-\mathbf{s}_{k}^{b}\|$ and $a_{t,(1)}=(\beta_{c,1},\ldots,\beta_{c,K})$, respectively. The reward of the first layer $r_{t,(1)}=(r_{t,1,(1)},\ldots,r_{t,K,(1)})$ can be calculated based on the communication overhead with the reward of UE $k$
\begin{equation}
\setcounter{equation}{20}
\begin{aligned}
r_{t,k,(1)}= \frac{1}{\sum_{m \in \mathcal{M}}|\mathbf{D}_{mk}|P_m^\mathrm{p,ap} + P_k^\mathrm{ue} + P_k^\mathrm{con}}.
\label{eq20}
\end{aligned}
\end{equation}

Due to the fact that maximizing EE performance in joint optimization problems is determined by both SE performance and power consumption, blindly reducing the power consumption solely on the reward function $r_{t,(1)}$ set at communication overhead is not appealing for improving EE performance. Therefore, to ensure that the generated actions can balance the two system indicators of system performance and communication overhead, we introduce a closed-form SE expression to optimize the original reward function $r_{t,k,(1)}$, which can be expressed as $r_{t,k,(1),n}= r_\mathrm{c} \cdot r_{t,k,(1)} + r_\mathrm{s}\cdot \text{SE}_{k,c}$,
where $r_\mathrm{c}$ and $r_\mathrm{s}$ denote the weight coefficients of communication overhead and SE performance in the updated reward function, respectively.

Essentially, the proposed DCP-MADDPG algorithm based on the MADDPG still adheres to the actor-critic architecture, which consists of the current critic network $\theta_{Q_{\pi_k},(l)}$ with an additional target critic network $\theta_{Q_{\pi'_k},(l)}$ responsible for updating policies and the current actor network $\theta_{{\pi_k},(l)}$ with an additional target actor network $\theta_{{\pi'_k},(l)}$ responsible for assigning actions, $\forall k \in \mathcal{K}$, $\forall l \in \{1, 2\}$. The communication process of the first layer is shown in Table \uppercase\expandafter{\romannumeral2}. Then, the policy gradient of the cooperative clustering network for $\pi_{k,(1)}$ can be modeled as
\begin{equation}
\setcounter{equation}{21}
\begin{aligned}
\nabla_{\theta_{\pi_k,(1)}}J(\theta_{\pi_k,(1)})=
{\mathbb{E}_{s_{t,(1)},a_{t,(1)}\sim \mathcal{D}_{(1)}}
\Big[\nabla_{\theta_{\pi_k,(1)}}\pi_k({a}_{t,k,(1)}|{s}_{t,k,(1)})
\nabla_{{\theta_{Q_{\pi_k},(1)}}}Q_{\theta_{Q_{\pi_k},(1)}}({s}_{t,(1)},{a}_{t,(1)})\Big]},
\label{eq21}
\end{aligned}
\end{equation}
where $\mathcal{D}_{(1)}$ denote the experience extraction replay buffer of the first layer that stores past experience $<s_{t,(1)}, a_{t,(1)},$ $ r_{t,(1),n}, s_{t+1,(1)}>$ \cite{[27]}.

Besides, considering that the value $Q_{\theta_{Q_{\pi_k}},(1)}({s}_{t,(1)},{a}_{t,(1)})$ is calculated by the current critic network, then the mean-squared Bellman error function \cite{[27]} of the critic network $L(\theta_{Q_{\pi_k},(1)})$ can be defined as
\begin{equation}
\setcounter{equation}{22}
\begin{split}
L(\theta_{Q_{\pi_k},(1)}) = \mathbb{E}_{s_{t,(1)},a_{t,(1)}\sim \mathcal{D}_{(1)}}\left[\left(Q_{\theta_{Q_{\pi_k},(1)}}({s}_{t,(1)},{a}_{t,(1)})-y_{t,k,(1)}\right)^2\right],
\label{eq22}
\end{split}
\end{equation}
where $y_{t,k,(1)}=r_{t,k,(1)}+\gamma_{(1)}(Q_{\theta_{Q_{\pi'_k},(1)}}({s}_{t,(1)}',{a}_{t,(1)}'))$ is the target value of agent $k$ with the target critic network value $Q_{\theta_{Q_{\pi'_k},(1)}}({s}_{t,(1)}',{a}_{t,(1)}')$.

Furthermore, in order to ensure that the target network $\theta_{Q_{\pi'_k},(l)}$ and $\theta_{{\pi'_k},(l)}$ remain stable, $\forall k \in \mathcal{K}$, the soft update is carried out with $\tau_{(1)}\ll 1$, and the target network of the first layer can be given by
\begin{equation}
\setcounter{equation}{23}
\begin{split}
\left \{
\begin{array}{ll}
{\theta_{\pi'_k,(1)}}\leftarrow\tau_{(1)}{\theta_{\pi'_k,(1)}}+(1-\tau_{(1)})\theta_{\pi_k,(1)},\\
{\theta_{Q_{\pi'_k},(1)}} \leftarrow \tau_{(1)}{\theta_{Q_{\pi'_k},(1)}} + (1-\tau_{(1)})\theta_{Q_{\pi_k},(1)}.
\end{array}
\right.
\label{eq23}
\end{split}
\end{equation}
\subsection{Second-Layer: Power Control Network}
\begin{table*}[t]
  \centering
  \fontsize{9}{7.5}\selectfont
  \caption{Communication Processed of MARL Network.}
  \label{Paper_comparison}
    \begin{tabular}{ !{\vrule width1.2pt}  m{1.8 cm}<{\centering} !{\vrule width1.2pt}   m{6.5 cm}<{\centering} !{\vrule width1.2pt}  m{6.5 cm}<{\centering} !{\vrule width1.2pt}}
    \Xhline{1.2pt}
        \rowcolor{gray!30} \bf Attributes & \bf First-Layer: Cooperative Clustering Network & \bf Second-Layer: Power Control Network \cr
    \Xhline{1.2pt}
         To whom?   & Broadcast the observed message to all agents & Multicast the observed message to all selected neighboring agents \cr\hline
         What?   & Observed coordinate information & Observed large-scale fading information \cr\hline
         How? & Combine coordinate information of all agents & Combine large-scale fading information of all selected neighboring agents \cr\hline
         Where? & UEs (coordinate information, scope of cooperative clustering) & UEs (large-scale fading information, moving information), APs (power information) \cr\hline
         When? & After all agents broadcast messages & After all agents receive neighboring messages \cr\hline
    \Xhline{1.2pt}
    \end{tabular}
\end{table*}
In contrast to the cooperative clustering network, the power control network still considers all UEs as independent agents deployed in dynamic scenarios, with the difference being that it maps the observed large-scale fading information to transmit power, moving step, and moving angle coefficients. Similarly, the communication process of the second layer is also shown in Table \uppercase\expandafter{\romannumeral2}.
It is worth noting that since the power control network is trained on the basis of the cooperative clustering network, each agent can exchange information with a selected subset of agents in the cooperative clustering network to complement their locally observed information.
Therefore, the corresponding observed state of the actor network and critic network at $t$ time slot can be defined as $s_{t,(2),a}=(s_{t,1,(2),a},\ldots,s_{t,K,(2),a})=(\beta_{1},\ldots,\beta_{K})$ with $\beta_{k}=\sum_{m \in \mathcal{M}}\beta_{mk}$ and $s_{t,(2),c}=(s_{t,1,(2),c},\ldots,s_{t,K,(2),c})$ with $s_{t,k,(2),c}=(s_{t,1,(2),a}{o}_{k,1},\ldots,s_{t,K,(2),a}{o}_{k,K})$, respectively.
Correspondingly, the assigned action of actor network and critic network can be defined as
$a_{t,(2),a}=(a_{t,(2),a}^p;a_{t,(2),a}^d;$ $a_{t,(2),a}^\theta)$ and $a_{t,(2),c}=(a_{t,(2),c}^p;a_{t,(2),c}^d;a_{t,(2),c}^\theta)$ with the power coefficient $a_{t,(2),c}^p=(a_{t,1,(2),a}^p{o}_{k,1},\ldots,$ $a_{t,K,(2),a}^p{o}_{k,K})$, moving step coefficient $a_{t,(2),c}^d=(a_{t,1,(2),a}^d{o}_{k,1},\ldots,a_{t,K,(2),a}^d{o}_{k,K})$, and moving angle coefficient
$a_{t,(2),c}^\theta=(a_{t,1,(2),a}^\theta{o}_{k,1},\ldots,a_{t,K,(2),a}^\theta{o}_{k,K})$, respectively.
Besides, the reward $r_{t,(2)}=(r_{t,1,(2)},\ldots,r_{t,K,(2)})$ can be calculated exploiting the updated closed-form EE expression with the reward $r_{t,k,(2)}(s_{t,k,(2),c},a_{t,k,(2),c})=\text{EE}_{k,c}$ at UE $k$.

Moreover, the policy gradient of the power control network for $\pi_{k,(2)}$ can be modeled as
\begin{equation}
\setcounter{equation}{24}
\begin{aligned}
\nabla_{\theta_{\pi_k,(2)}}J(\theta_{\pi_k,(2)})=
{\mathbb{E}_{s_{t,(2)},a_{t,(2)}\sim \mathcal{D}_{(2)}}
\Big[\nabla_{\theta_{\pi_k,(2)}}\pi_k({a}_{t,k,(2),a})
\nabla_{{\theta_{Q_{\pi_k},(2)}}}Q_{\theta_{Q_{\pi_k},(2)}}({s}_{t,k,(2),c},{a}_{t,k,(2),c})\Big]},
\label{eq26}
\end{aligned}
\end{equation}
where $\mathcal{D}_{(2)}$ denote the experience extraction replay buffer of the second layer that stores past experience $<s_{t,(2),c}, a_{t,(2)},$ $ r_{t,(2)}, s_{t+1,(2),c}>$ \cite{[27]}.

Similarly, the mean-squared Bellman error function of the current critic network $L(\theta_{Q_{\pi_k},(2)})$ can be defined as
\begin{equation}
\setcounter{equation}{25}
\begin{split}
L(\theta_{Q_{\pi_k},(2)}) = \mathbb{E}_{s_{t,(2),c},a_{t,(2)}\sim \mathcal{D}_{(2)}}\bigg[\Big(Q_{\theta_{Q_{\pi_k},(2)}}({s}_{t,k,(2),c},{a}_{t,k,(2),c})-y_{t,k,(2)}\Big)^2\bigg],
\label{eq27}
\end{split}
\end{equation}
where the action value $Q_{\theta_{Q_{\pi_k}},(2)}({s}_{t,k,(2),c},{a}_{t,k,(2),c})$ is calculated by the current critic network $\theta_{Q_{\pi_k},(2)}$, and $y_{t,k,(2)}=r_{t,k,(2)}+\gamma_{(2)}(Q_{\theta_{Q_{\pi'_k},(2)}}({s}_{t,k,(2),c}',{a}_{t,k,(2),c}'))$ is the target value with the target critic network value $Q_{\theta_{Q_{\pi'_k},(2)}}({s}_{t,k,(2),c}',{a}_{t,k,(2),c}')$.

Furthermore, the soft update is carried out with $\tau_{(2)}\ll 1$ to ensure that the target network $\theta_{Q_{\pi'_k},(2)}$ and $\theta_{{\pi'_k},(2)}$ remain stable, $\forall k \in \mathcal{K}$, which are given by
\begin{equation}
\setcounter{equation}{26}
\begin{split}
\left \{
\begin{array}{ll}
    {\theta_{\pi'_k,(2)}}\leftarrow\tau_{(2)}{\theta_{\pi'_k,(2)}}+(1-\tau_{(2)})\theta_{\pi_k,(2)},\\
    {\theta_{Q_{\pi'_k},(2)}} \leftarrow \tau_{(2)}{\theta_{Q_{\pi'_k},(2)}} + (1-\tau_{(2)})\theta_{Q_{\pi_k},(2)}.
\end{array}
\right.
\label{eq26}
\end{split}
\end{equation}
\begin{algorithm}[t]
  \caption{Dynamic Clustering-Power MADDPG}
  \label{alg::conjugateGradient}
  \begin{algorithmic}[1]
      \State
        \textbf{Initialize} Observations of the first layer ${s}_{t,1,(1)},\ldots,$ ${s}_{t,K,(1)}$ and the second layer ${s}_{t,1,(2),a},\ldots, {s}_{t,K,(2),a}$
      \For {episode = 1 to max-episodes}
        \For {step = 1 to max-steps}
            \State Actor network determines the action of the first and second layer: ${a}_{t,k,(1)}$ = $\pi_k$(${s}_{t,k,(1)}$), ${a}_{t,k,(2)}$ = $\pi_k$(${s}_{t,k,(2),a}$)
            \State Update the indicator function $\mathbf{o}_k$ based on ${a}_{t,k,(1)}$
            \State Update the observed state of critic network ${s}_{t,k,(2),c}$ based on $\mathbf{o}_k$ and ${s}_{t,k,(2),a}$
            \State Obtain the expected rewards of the first layer $r_{t,k,(1),n}$ and the second layer $r_{t,k,(2)}$
            \State Update the action of critic network ${a}_{t,k,(2),c}$
            \State Obtain the next state $s_{t+1,k,(1)}$, $s_{t+1,k,(2),a}$, and ${s}_{t+1,k,(2),c}$ after the agent interacts with the environment
            \State Store experience $<{s}_{t,(1)}, {a}_{t,(1)},{r}_{t,(1)}, {s}_{t+1,(1)}>$ to $\mathcal{D}_{(1)}$ and $<{s}_{t,(2),c}, {a}_{t,(2),c}, $ ${r}_{t,(2)},$ ${s}_{t+1,(2),c}>$ to $\mathcal{D}_{(2)}$
            \If {update the network $l \in \{1,2\}$}
              \For {agent $k$ = 1 to $K$}
                \State Sample a mini-batch $\mathcal{B}_{k,(l)}$ from $\mathcal{D}_{(l)}$
                \State Calculate the critic value $L(Q_{\theta_{Q_{\pi,k}},(l)})$
                \State Update the weights of critic network
                \State Calculate the policy gradient
                \State Update the target network ${\theta_{{\pi_k'},(l)}}$, ${\theta_{Q_{\pi_k'},(l)}}$
              \EndFor
            \EndIf
        \EndFor
      \EndFor
  \end{algorithmic}
\end{algorithm}
\begin{table}[t]
\centering
    \fontsize{9}{7}\selectfont
    \caption{The Model Structure and Experimental Details.}
    \label{paper}
    \begin{tabular}{ccc}
    \toprule
    \bf Parameters &  \bf Size \\
    \midrule
    1st hidden layer & 128, Leaky Relu (0.01)\\
    2nd hidden layer & 64, Leaky Relu (0.01) \\
    Discounted factor $\gamma_{(1)}$ and $\gamma_{(2)}$ & 0.99 and 0.99 \\
    Experience pool size $\mathcal{D}_{(1)}$ and $\mathcal{D}_{(2)}$ & 128 and 256 \\
    Maximal gradient & 0.5 \\
    Soft update rate $\tau_{(1)}$ and $\tau_{(2)}$ & 0.01 and 0.01 \\
    Reward weight $r_\mathrm{c}$ and $r_\mathrm{s}$ & 0.75 and 0.25\\
    Random episodes & 2000\\
    \bottomrule
    \end{tabular}
\end{table}

The procedure of DCP-MADDPG for maximizing EE is summarized in $\textbf{Algorithm 1}$.
\section{Numerical Results}
In this section, we evaluate the performance of the joint dynamic clustering and power control scheme with different parameters and consider a CF XL-MIMO system, where all APs and UEs are uniformly distributed in a square area of size $1 \times 1$ $\text{km}^2$, which is wrapped around at the edges to avoid boundary effects \cite{[8]}.
Then, we consider a free-space path loss model for the considered CF XL-MIMO system \cite{[4]} and denote the upper limit of the transmission power of each UE as $p_{k}=$ 200 mW, $\forall k \in \mathcal{K}$. Moreover, the model structure and experimental details of the first and second layer are shown in Table \uppercase\expandafter{\romannumeral3}, and the simulation works are conducted with an Nvidia GeForce GTX 3060 Graphics Processing Unit.
\subsection{Comparative Analysis of Computational Complexity}
In this subsection, we compare the computational complexity of different network architectures, which consist of two parts: cooperative clustering network and MARL-based neural network.
For the proposed DCP-MADDPG, the computational complexity of the cooperative clustering network is $\mathcal{O}(N_\mathrm{B}N_\mathrm{P}K)$, where $N_\mathrm{B}$ and $N_\mathrm{D}$ are the average number of APs served by each UE and the average number of UEs served by the same AP for information sharing, respectively. Moreover, the computational complexity of MARL-based neural network is $\mathcal{O}((K^3d_{a,(1)}+N_\mathrm{B}K^2d_{a,(2)})\sum_{a=1}^{A}Q_a^{2}+ (K^4+K^2d_{a,(1)} +N_\mathrm{P}N_\mathrm{B}K^2+N_\mathrm{P}Kd_{a,(2)})\sum_{c=1}^{C}Q_c^{2})$, where $C$ and $A$ are the number of hidden layers for the critic network and actor network with the dimension of the output action $d_{a,(1)}$ and $d_{a,(2)}$, respectively, and $Q_a$ denotes the output size of the $a$-th layer or the input size of the next layer. As for the remaining algorithms, i.e., K-Means, MUEGA, R-Variable, and Q-Variable, which differ from the proposed algorithm, the network architecture of these algorithms only consists of a single-layer power control network, resulting in differences in computational complexity of the dynamic clustering and the MARL-based neural network, as shown in Table \uppercase\expandafter{\romannumeral4}, where $N_\mathrm{K}$, $N_\mathrm{M}$, $N_\mathrm{R}$, and $N_\mathrm{Q}$ denote the average number of UEs served by the same AP for information sharing under K-Means, MUEGA, R-Variable, and Q-Variable, respectively. Moreover, Table \uppercase\expandafter{\romannumeral4} also presents a comparison of running time over different network architectures mentioned earlier. We can clearly observe that compared to various network architectures with information sharing, the proposed DCP-MADDPG has a significant performance improvement in terms of running time. For example, compared to the CTDE mechanism and fully centralized architecture, the runtime is reduced by 53.31\% and 79.96\%, respectively.
\begin{table*}[t]
  \centering
  \fontsize{9}{7}\selectfont
  \caption{Comparison of Computational Complexity and Running Time over Different Network Architectures}
  \label{Paper_comparison}
    \begin{tabular}{ !{\vrule width1.2pt}  m{1.8 cm}<{\centering} !{\vrule width1.2pt}  m{3.25 cm}<{\centering} !{\vrule width1.2pt} m{7 cm}<{\centering} !{\vrule width1.2pt} m{2.5 cm}<{\centering} !{\vrule width1.2pt}}
    \Xhline{1.2pt}
        \rowcolor{gray!30} \bf Methods  &  \bf Cooperative Clustering &  \bf MARL-based Neural Network &  \bf Running Time [s] \cr
    \Xhline{1.2pt}
        Decentralized & $\mathcal{O}(N_\mathrm{B}K)$ & $\mathcal{O}(N_\mathrm{B}K^{2}d_{a}\sum_{a=1}^{A}Q_a^{2}+(N_\mathrm{B}K^{2}+Kd_a)\sum_{c=1}^{C}Q_c^{2})$ & 0.75 (-9.64\%) \cr\hline
         CTDE   & $\mathcal{O}(MK^2)$  & $\mathcal{O}(MK^{2}d_{a}\sum_{a=1}^{A}Q_a^{2} + (MK^{3}+K^2d_a)\sum_{c=1}^{C}Q_c^{2})$ & 1.78 (+114.46\%) \cr\hline
         Centralized & $\mathcal{O}(MK^2)$  & $\mathcal{O}(MK^{3}d_{a}\sum_{a=1}^{A}Q_a^{2} + (MK^{3}+K^2d_a)\sum_{c=1}^{C}Q_c^{2})$ & 4.15 (+400.01\%) \cr\hline
         K-Means  & $\mathcal{O}(N_\mathrm{K}K^2+N_\mathrm{K}N_\mathrm{B}K)$  & $\mathcal{O}(N_\mathrm{B}K^2d_a\sum_{a=1}^{A}Q_a^{2}+ (N_\mathrm{K}N_\mathrm{B}K^2+N_\mathrm{K}Kd_a)\sum_{c=1}^{C}Q_c^{2})$ & 0.97 (+16.87) \cr\hline
         MUEGA  & $\mathcal{O}(K^3+N_\mathrm{M}N_\mathrm{B}K)$ & $\mathcal{O}(N_\mathrm{B}K^2d_a\sum_{a=1}^{A}Q_a^{2}+ (N_\mathrm{M}N_\mathrm{B}K^2+N_\mathrm{M}Kd_a)\sum_{c=1}^{C}Q_c^{2})$ & 0.91 (+9.64\%) \cr\hline
         R-Variable  & $\mathcal{O}(K^2+N_\mathrm{R}N_\mathrm{B}K)$ & $\mathcal{O}(N_\mathrm{B}K^2d_a\sum_{a=1}^{A}Q_a^{2}+ (N_\mathrm{R}N_\mathrm{B}K^2+N_\mathrm{R}Kd_a)\sum_{c=1}^{C}Q_c^{2})$ & 0.85 (+2.41\%)  \cr\hline
         Q-Variable & $\mathcal{O}(N_\mathrm{Q}^2K^2+N_\mathrm{Q}N_\mathrm{B}K)$ & $\mathcal{O}(N_\mathrm{B}K^2d_a\sum_{a=1}^{A}Q_a^{2}+ (N_\mathrm{Q}N_\mathrm{B}K^2+N_\mathrm{Q}Kd_a)\sum_{c=1}^{C}Q_c^{2})$ & 0.88 (+6.02\%) \cr\hline
         \bf Proposed  & $\mathcal{O}(N_\mathrm{P}K^2+N_\mathrm{P}N_\mathrm{B}K)$ & $\mathcal{O}((K^3d_{a,(1)}+N_\mathrm{B}K^2d_{a,(2)})\sum_{a=1}^{A}Q_a^{2}+ (K^4+K^2d_{a,(1)} +N_\mathrm{P}N_\mathrm{B}K^2+N_\mathrm{P}Kd_{a,(2)})\sum_{c=1}^{C}Q_c^{2})$ & 0.83 (Proposed) \cr\hline
    \Xhline{1.2pt}
    \end{tabular}
  \vspace{0cm}
\end{table*}
\subsection{Effects of the Number of UEs Participating in Information Sharing}
We investigate the impact of different numbers of UEs participating in information sharing on the proposed DCP-MADDPG and conventional algorithms with different network architectures. Fig. 3 and Fig. 4 show the average achievable sum SE and EE performance against the number of UEs participating in information sharing, respectively.
In Fig. 3, numerical results show that as the number of UEs participating in information sharing increases, each UE receives more state information that facilitates the design of effective cooperative clustering architectures. In particular, with a sufficient number of UEs participating in information sharing, the SE performance of different cooperative clustering architectures approaches that of conventional centralized architectures. Similarly, in Fig. 4, as the number of UEs participating in information sharing gradually increases, the gains in SE performance are outweighed by the escalated total power consumption, leading to a gradual decrease in EE performance under various algorithms of different network architectures to the conventional CTDE mechanism.
This phenomenon indicates that there exists a non-trivial trade-off between the number of UEs participating in information sharing and EE performance. In particular, an appropriate number of UEs participating in information sharing is crucial for balancing SE and EE performance.

In addition, we compare two dynamic clustering architectures based on different reward function forms. Our observations indicate that a well-balanced reward function that accounts for both SE and communication power consumption significantly enhances system performance. For example, compared to the reward function scheme designed solely based on communication power consumption, the reward function scheme combining SE expression yields a 2.77\% and 11.76\% improvement in SE and EE, respectively.
\begin{figure}[t]
	\centering
	\begin{minipage}[t]{0.48\textwidth}
	\centering
    \includegraphics[scale=0.5]{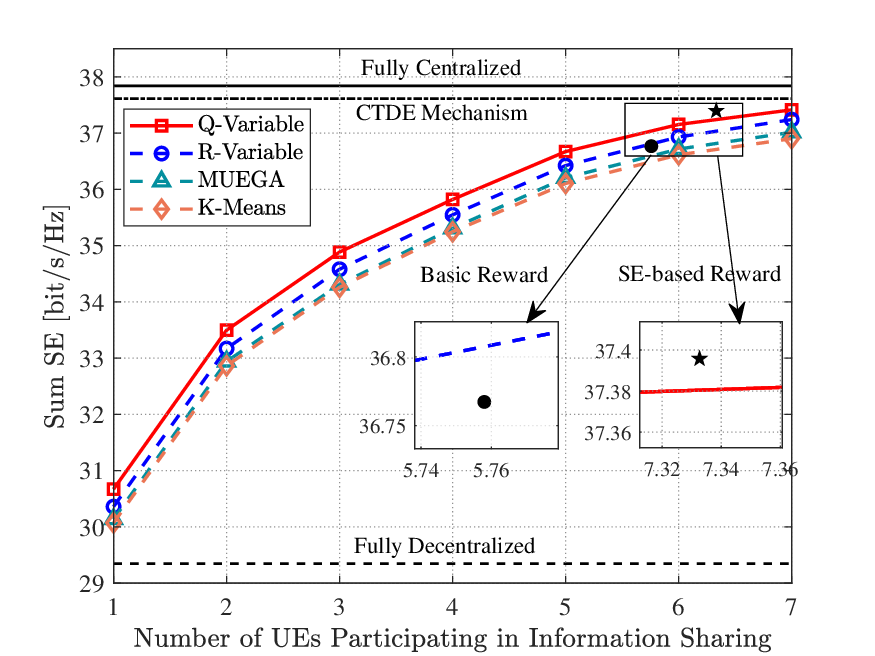}
    \caption{The average achievable sum SE versus the number of UEs participating in information sharing for MR combining with $M=9$, $K=6$, $N_r=N_{H_r}\times N_{V_r}=81$, $N_s=N_{H_s}\times N_{V_s}=9$, and $\Delta_r = \lambda/3$.}
	\end{minipage}
    \hfill%
	\begin{minipage}[t]{0.48\textwidth}
	\centering
    \includegraphics[scale=0.5]{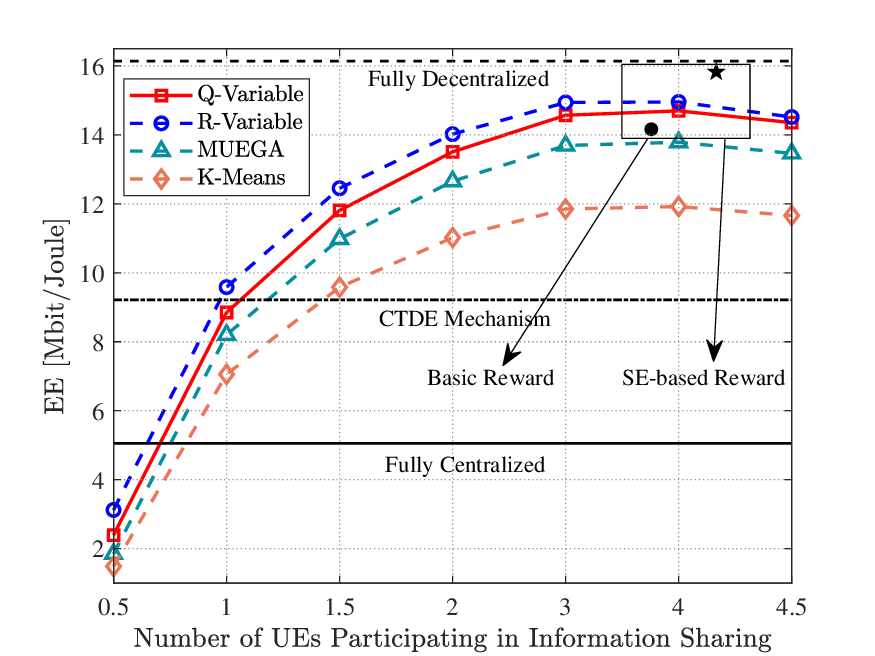}
    \caption{The average achievable EE versus the number of UEs participating in information sharing for MR combining with $M=9$, $K=6$, $N_r=N_{H_r}\times N_{V_r}=81$, $N_s=N_{H_s}\times N_{V_s}=9$, and $\Delta_r = \lambda/3$.}
	\end{minipage}
\end{figure}
\subsection{Effects of the Number of UEs and BSs}
We investigate the impact of different system parameters on the proposed DCP-MADDPG and conventional algorithms with different network architectures. Fig. 5 and Fig. 6 show the average achievable EE and communication power consumption against the number of UEs, respectively. In Fig. 5, the numerical results show that the EE performance of various algorithms, each with different network architecture increases as the number of UEs increases. In particular, the proposed DCP-MADDPG has a significant advantage in improving EE performance. Compared with the conventional CTDE mechanism, the performance gap in EE between it and the proposed DCP-MADDPG gradually widens as the number of UEs increases. For instance, the performance gaps are 28.46\% and 55.44\% for $K=4$ and $K=9$, respectively. The reason for the increase in this performance gap is that the cooperative clustering architecture helps to reduce redundant communication among UEs in the proposed DCP-MADDPG scheme, thereby avoiding unnecessary communication overhead.
In Fig. 6, we can observe that as the number of UEs increases, the communication quantity and communication overhead of each UE will increase accordingly, while the number of APs served by each UE remains almost unchanged. In addition, the proposed DCP-MADDPG significantly reduces the communication overhead by selecting appropriate AP clusters and neighboring UEs, for example, by 66.38\% compared to the conventional centralized schemes at the number of UEs $K=9$. This phenomenon indicates that the proposed DCP-MADDPG is effective in utilizing the system resources for improving EE performance, compared with the conventional centralized schemes.
\begin{figure}[t]
	\centering
	\begin{minipage}[t]{0.48\textwidth}
	\centering
    \includegraphics[scale=0.5]{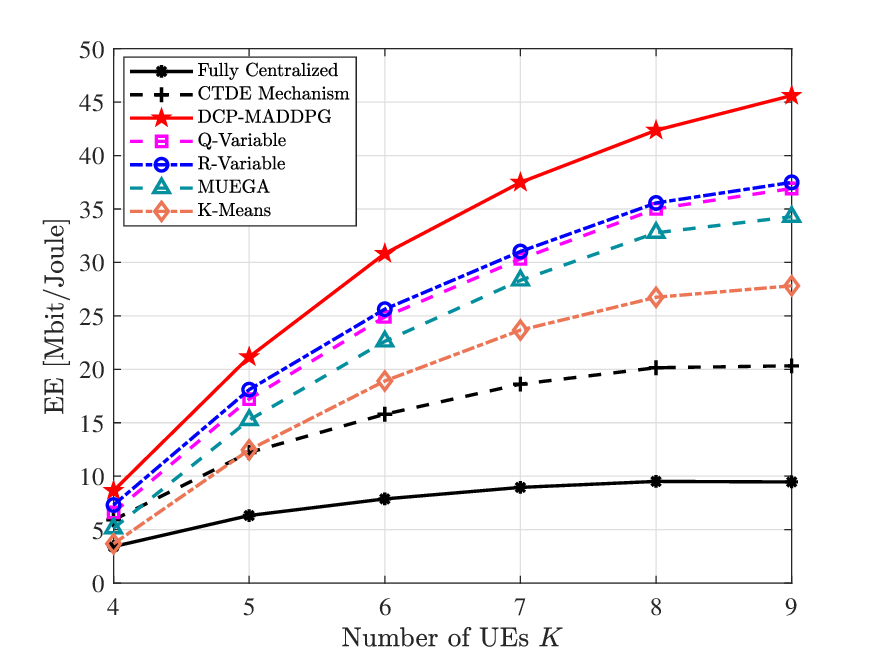}
    \caption{The average achievable EE versus the number of UEs for MR combining with $M=16$, $N_r=N_{H_r}\times N_{V_r}=81$, $N_s=N_{H_s}\times N_{V_s}=9$, and $\Delta_r = \lambda/3$.}
	\end{minipage}
    \hfill%
	\begin{minipage}[t]{0.48\textwidth}
	\centering
    \includegraphics[scale=0.5]{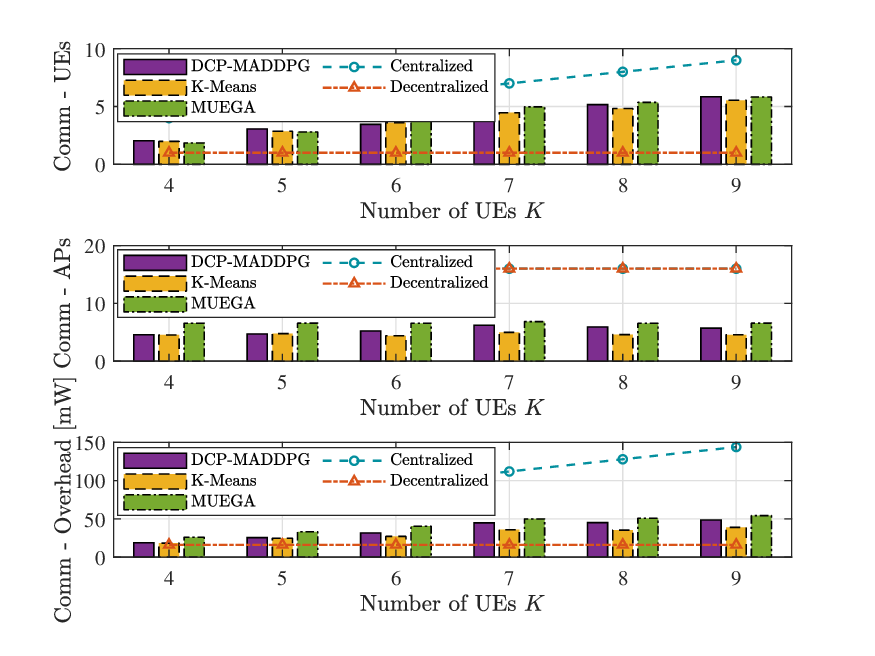}
    \caption{The average communication consumption versus the number of UEs for MR combining with $M=16$, $N_r=N_{H_r}\times N_{V_r}=81$, $N_s=N_{H_s}\times N_{V_s}=9$, and $\Delta_r = \lambda/3$.}
	\end{minipage}
\end{figure}
\begin{figure}[t]
	\centering
	\begin{minipage}[t]{0.48\textwidth}
	\centering
    \includegraphics[scale=0.5]{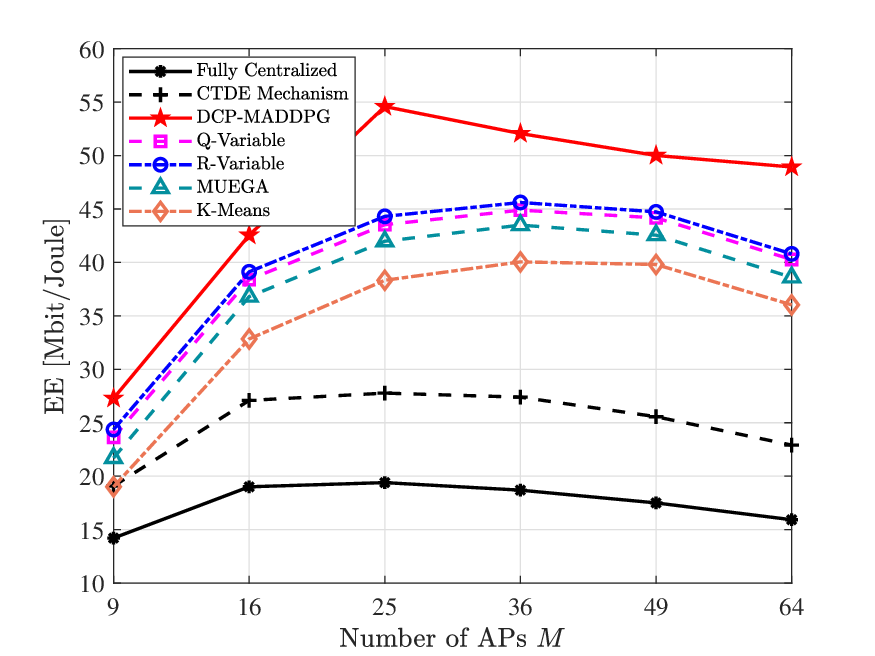}
    \caption{The average achievable EE versus the number of APs for MR combining with $K=8$, $N_r=N_{H_r}\times N_{V_r}=81$, $N_s=N_{H_s}\times N_{V_s}=9$, and $\Delta_r = \lambda/3$.}
	\end{minipage}
    \hfill%
	\begin{minipage}[t]{0.48\textwidth}
	\centering
    \includegraphics[scale=0.5]{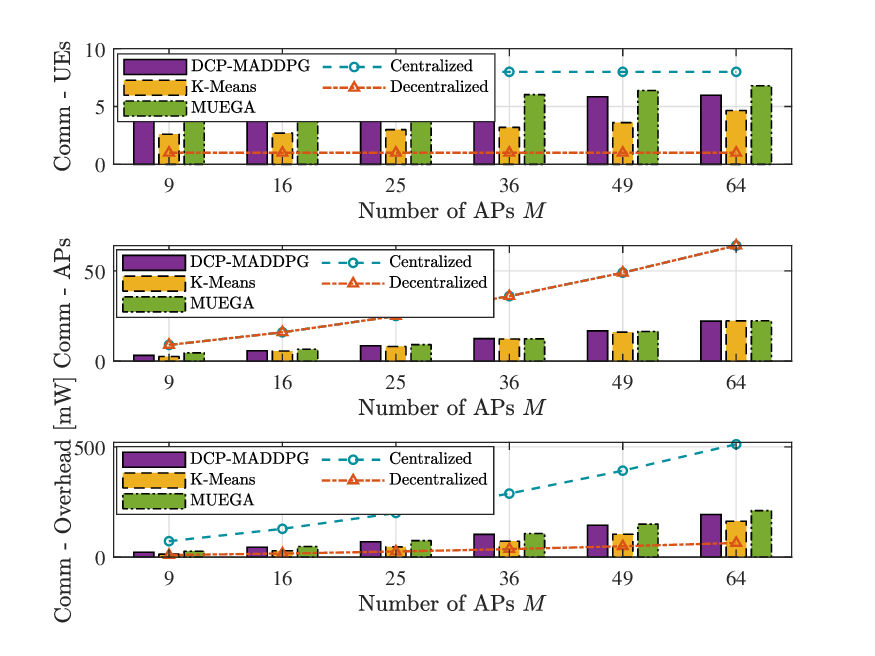}
    \caption{The average communication consumption versus the number of APs for MR combining with $K=8$, $N_r=N_{H_r}\times N_{V_r}=81$, $N_s=N_{H_s}\times N_{V_s}=9$, and $\Delta_r = \lambda/3$.}
	\end{minipage}
\end{figure}

Similarly, Fig. 7 and Fig. 8 depict the average achievable EE and communication power consumption against the number of BSs, respectively.
In Fig. 7, the performance gap in EE performance between the conventional CTDE mechanism and the proposed DCP-MADDPG still increases as the number of APs increases, e.g. the performance gaps are 30.08\% and 53.17\% over $M=9$ and $M=64$, respectively. This indicates that by selecting appropriate AP clusters, the cooperative cluster architecture can significantly reduce communication power consumption without sacrificing too much SE performance, thereby effectively improving EE performance.
For example, in Fig. 8, as the number of APs increases from $M=9$ to $M=64$, the proposed DCP-MADPG consistently reduces communication overhead by more than 62.24\% compared to the conventional centralized scheme.
\subsection{Effects of the Number of Antennas Per UE}
Next, to further highlight the advantages of the proposed DCP-MADDPG, we investigate the impact of different numbers of antennas per UE on the proposed DCP-MADDPG and conventional algorithms with different network architectures.
The average achievable EE as a function of the number of antennas per UE with MR combining over different schemes is shown in Fig. 9. It can be observed that the performance gap in EE performance between the conventional architectures and the proposed DCP-MADDPG increases as the number of antennas per UE increases, e.g. the performance gaps are 44.27\% and 66.74\% for $N_s=9$ and $N_s=64$, respectively. Moreover, we can observe that there are diminishing returns when the number of antennas is sufficiently large, e.g. $N_s \geqslant 49$. The reason is that the power consumption loss generated by the increased number of antennas neutralize the SE performance gain. Therefore, the EE can greatly benefit from equipping multiple UE antennas for improving communication quantity, thereby effectively reducing the required power consumption without significantly compromising SE performance.
\begin{figure}[t]
	\centering
	\begin{minipage}[t]{0.48\textwidth}
	\centering
    \includegraphics[scale=0.5]{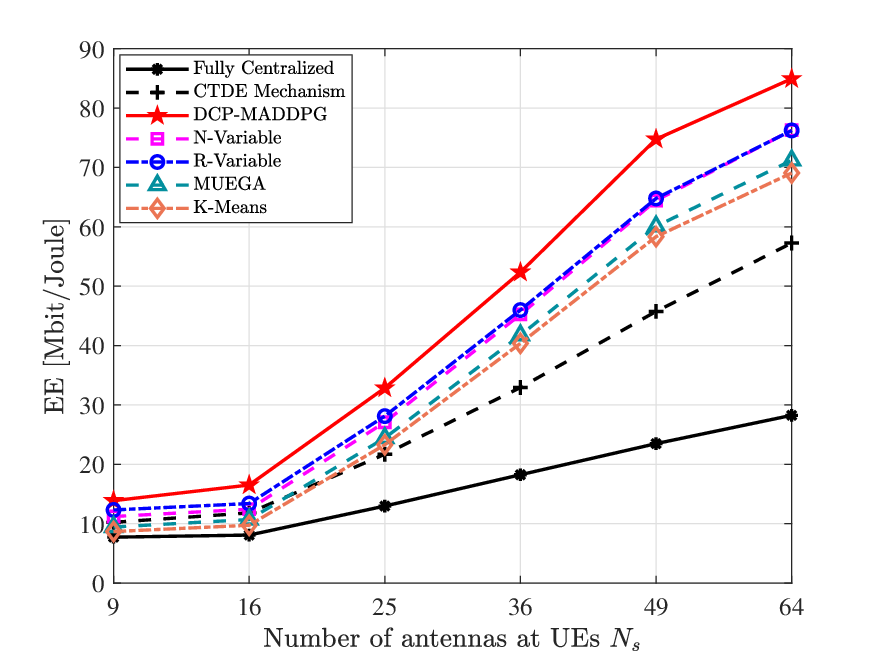}
    \caption{The average achievable EE versus the number of antennas at UEs for MR combining with $M=16$, $N_r=N_{H_r}\times N_{V_r}=81$, $N_s=N_{H_s}\times N_{V_s}=9$, and $\Delta_r = \lambda/3$.}
	\end{minipage}
    \hfill%
	\begin{minipage}[t]{0.48\textwidth}
	\centering
    \includegraphics[scale=0.5]{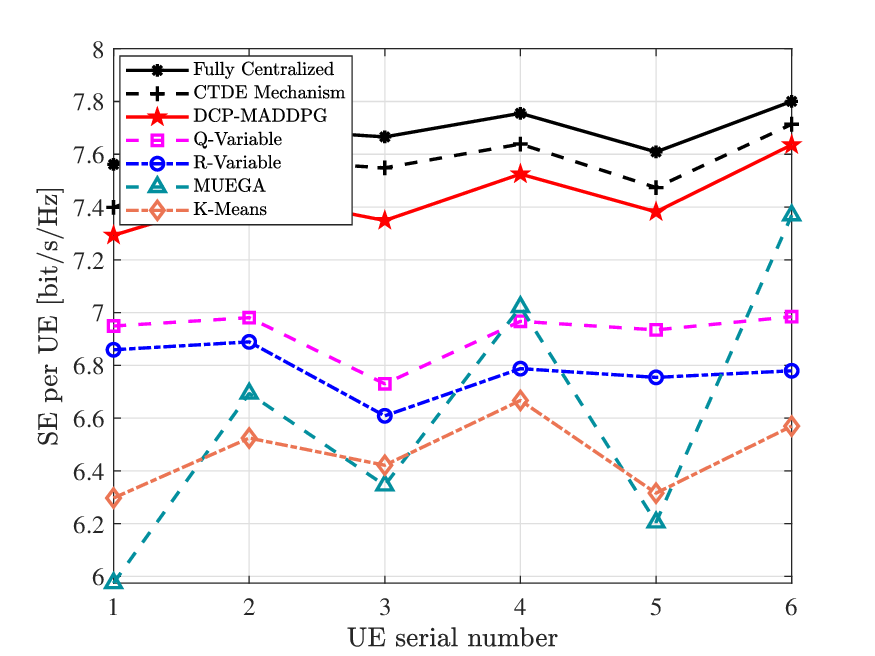}
    \caption{The achievable SE at each  UEs for MR combining with $M=16$, $K=6$, $N_r=N_{H_r}\times N_{V_r}=81$, $N_s=N_{H_s}\times N_{V_s}=9$, and $\Delta_r = \lambda/3$.}
	\end{minipage}
\end{figure}
\begin{figure}[t]
\centering
    \includegraphics[scale=0.5]{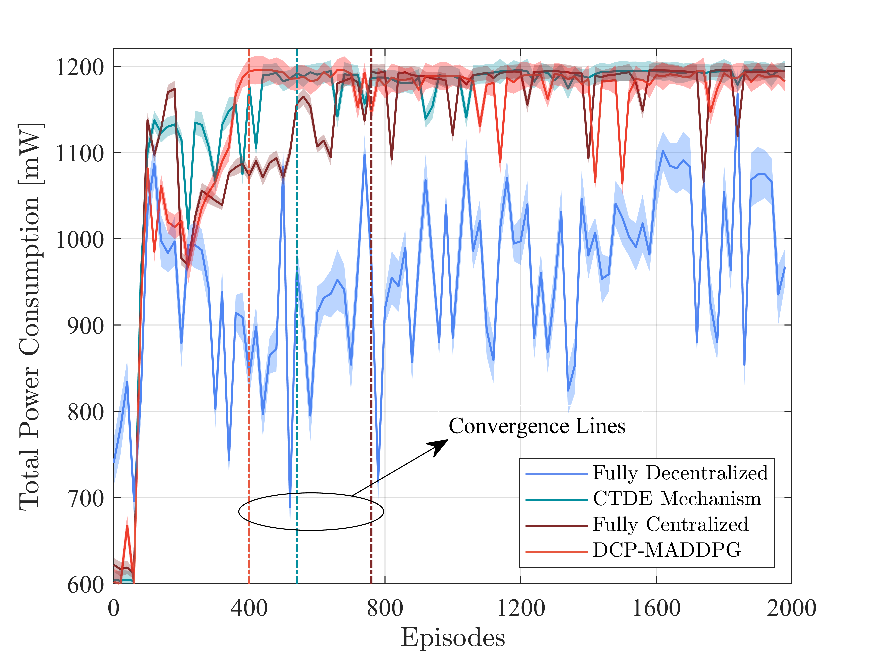}
    \caption{Convergence rate over different network architectures for MR combining with $M=9$, $K=6$, $N_r=N_{H_r}\times N_{V_r}=81$, $N_s=N_{H_s}\times N_{V_s}=9$, and $\Delta_r = \lambda/3$.
    \label{fig44}}
\end{figure}
\subsection{Effects of AP Selection on the Performance of Each UE}
Moreover, considering that one of the main themes in cell-free XL-MIMO architectures is to provide unified performance for all UEs within the system, we investigate the impact of AP selection schemes designed under different network architectures on the performance of each UE.
The achievable SE for each UE with MR combining on different optimization schemes is shown in Fig. 10. It can be observed that the performance fluctuation between different UEs in the proposed DCP-MADDPG algorithm is within 4.48\%, which is close to the performance fluctuation of 4.08\% - 4.81\% under two fully connected schemes including fully centralized and CTDE mechanism. This indicates that the designed AP selection scheme satisfies the requirement of providing unified performance for all UEs in cell-free XL-MIMO systems. The reason is that the first layer of the cooperative clustering network can allocate a portion of APs outside the neighborhood to each UE for service based on the initial AP selection scheme, thereby achieving relatively consistent performance.
In contrast, conventional MUEGA-based optimization architectures exhibit significant performance fluctuations between UEs under their designed AP selection scheme, e.g., 18.92\%, which is much greater than the performance fluctuations under conventional fully connected schemes.
This results in the inability to provide similar performance for each UE. Therefore, a suitable cooperative clustering network can better achieve a balance between system performance and communication overhead while ensuring fair performance among UEs.
\subsection{Comparison of Convergence Rate}
In this subsection, we investigate the convergence rate over various algorithms with different network architectures in mobile scenarios for MR combining.
Fig. 11 shows a comparison of the convergence rate with $M=9$, $K=6$, $N_r=N_{H_r}\times N_{V_r}=81$, $N_s=N_{H_s}\times N_{V_s}=9$, and $\Delta_r = \lambda/3$. Compared with centralized architectures, e.g. fully centralized and CTDE mechanism, the proposed DCP-MADDPG improves the convergence rate by reducing the required communication overhead among redundant agents, respectively increasing 25.93\% and 47.37\%. Additionally, compared with decentralized architectures, our proposed DCP-MADDPG selects only neighboring agents served by the same AP for information sharing, avoiding non-convergence issues similar to those in decentralized architectures.
\section{Conclusion}
In this paper, we focused on a user-centric mobile CF XL-MIMO system, where cooperative clustering ensures that each UE is only served by a set of neighboring APs and only share information with neighboring UEs served by the same AP. Then, we proposed a novel double-layer MARL scheme that addresses the joint resource allocation problem consisting of cooperative clustering and power control design, where all clusters are adaptively optimized. The simulation results verified that the designed double-layer MARL scheme can effectively strike an excellent balance between system performance and communication overhead, thereby achieving a significant enhancement in EE while approaching SE under conventional centralized optimization. In future work, we will integrate MARL with emergent communication such that we can dynamically adjust the communication strategy by establishing systematic and structured communication protocols, as well as to explore the impact of different communication protocols on system performance.
\bibliographystyle{IEEEtran}
\bibliography{IEEEabrv,Ref}
\end{document}